\documentclass[11pt,a4paper]{article}
\usepackage{jcappub,bookmark}
\usepackage{amsmath,amssymb,graphics,epsfig,subfigure,color}
\usepackage{multirow}
\newcommand {\nn}    {\nonumber}

\title{Localization and mass spectra of various matter fields on scalar-tensor brane}

\author{Qun-Ying Xie$^{a,b,}$\footnote{Corresponding author},
        Zhen-Hua Zhao$^{c}$,
        Yi Zhong$^{b}$,
        Jie Yang$^{b}$,
        Xiang-Nan Zhou$^{b}$}
\affiliation{
  $^{a}$School of Information Science and Engineering, Lanzhou University, Lanzhou 730000, People's Republic of China\\
  $^{b}$Institute of Theoretical Physics, Lanzhou University, Lanzhou 730000, People's Republic of China\\
  $^{c}$Department of Applied Physics, Shandong University of Science and Technology, Qingdao, 266590, People's Republic of China
          }

\emailAdd{xieqy@lzu.edu.cn}

\abstract{
Recently, a new scalar-tensor braneworld model was presented in [Phys. Rev. \textbf{D 86} (2012) 127502]. It
not only solves the gauge hierarchy problem but also reproduces a correct Friedmann-like
equation on the brane.
In this new model, there are two different brane solutions,
for which the mass spectra of gravity on the brane are the same.
In this paper, we investigate
localization and mass spectra of various
bulk matter fields (i.e., scalar, vector, Kalb-Ramond, and fermion fields)
on the brane. It is shown that the zero modes of all the matter fields can be localized on the positive tension brane under some conditions, and the mass spectra of each kind of bulk matter field for the two brane solutions are different,
which implies that the two brane solutions are not physically equivalent.
}

\keywords{Large Extra Dimensions, Field Theories in Higher Dimensions}

\begin{document}

\maketitle

\section{Introduction}

Localization of gravity and various bulk matter fields on a braneworld is always an interesting and important issue for braneworld models. In the Randall-Sundrum (RS) braneworld model \cite{RSmodel1,RSmodel2} and its generations (see Refs.~\cite{Hatanaka1999,Das2008,Lepe2008,Zhong2010Scalar-Kinetic,Yang2010Weyl,
Balcerzak20011,Ahmed20014,Aros2013,Fu1407.6107,WangBin2014,Bazeia1411.0897} for examples), the extra dimension can be finite or infinite and its geometry is warped. In order to recover the familiar four-dimensional Newtonian potential, the zero mode of gravity should be localized on the brane. While the massive Kaluza-klein (KK) modes of gravity will give modification to the Newtonian potential. This modification is very different from the case of the Arkani-Hamed-Dimopoulos-Dvali (ADD) braneworld model \cite{Arkani-Hamed1998}. On the other hand,
the matters are either assumed to be confined on the brane and hence are four dimensional \cite{RSmodel1,RSmodel2},
or are assumed to be bulk fields propagating in the five-dimensional space-time.
For the later case,
one needs some localization mechanisms to trap at least the zero modes of various matter fields on the brane because the effective physics in low energy scale is four dimensional.

It is known that a free massless scalar field
can be localized on the RS brane and its generalized branes \cite{Bajc2000,Fuchune,Oda2000}.
While a free vector field cannot be localized on the RS brane. But it can be trapped on the generalized RS branes with codimension two or more \cite{Oda2000}, or on the dS branes
and the Weyl branes \cite{Liu0708,LiuPRD2008}, or on the brane generated by two scalar fields \cite{Fuchune}.
Localization of the antisymmetric Kalb-Ramond (KR) tensor field is similar to that of the vector field \cite{Fuchune}. If the couplings with background fields are introduced, then localization and mass spectra of the vector and KR fields will be more interesting and complicated, see Refs. \cite{Massimo2002,ZhaoZHH2014,Carlos2014,Cruz2009,Christiansen2010,Christ2012,CruzWT,DuYunZhi} for details.
Localization of the spin 1/2 fermion on the brane is
extremely important.
In many previous papers \cite{Oda2000,Volkas2007,RandjbarPLB2000,GuoHeng2013}, it has been proved that
in order to localize the massless fermion on the brane, one should
introduce a localization
mechanism such as Yukawa coupling.

In the RS1 model \cite{RSmodel1}, there are two branes that locate at the boundaries of a compact extra dimension with topology $S^1/Z_2$. One is the negative tension brane (called Tev brane) and another is the positive tension brane (called {Planck} brane).
Matters and gravity are localized on the Tev and {Planck} branes, respectively. Thus, our Universe is resided on the Tev brane. The famous gauge hierarchy problem is solved in this model due to the warping of the extra dimension.
However, there exists a severe
cosmological problem in RS1 model because it will lead to a ``wrong-signed" Friedmann-like equation on our Universe. This problem can be avoided if our Universe is confined on the positive tension brane \cite{Csaki1999,Cline1999,Shiromizu2000}.
Recently, the authors in Ref. \cite{YangKe} investigated a simple
generation of the RS1 model in the scalar-tensor
gravity by changing the profile of the
massless graviton to move our world to the positive tension
brane. The scalar-tensor
brane model can solve not only the cosmological problem
in the RS1 model but also the gauge hierarchy problem.
In this model, there are two similar but different brane solutions.
The zero modes and mass spectra of gravity for both brane solutions are the same.
Therefore, the two solutions cannot be distinguished from the mass spectra of gravity.

In this paper, in order to distinguish the two brane solutions of
the scalar-tensor brane proposed in Ref. \cite{YangKe} and show their rich structure,
we would like to study localization and mass spectra of various
bulk matter fields (i.e., scalar, vector, KR, and fermion fields) on the scalar-tensor brane for
the two brane solutions. We will introduce the couplings between the matters and the background scalar
(i.e., the dilaton that generates the brane).
It will be shown that these couplings are necessary for most of the cases we consider.
Especially, for the case of fermion, the usual Yukawa coupling {$\eta \bar \Psi F(\phi)\Psi$}
does not work anymore because the background scalar has even parity instead of odd one,
so we adopt the new scalar-fermion coupling $\eta \bar \Psi \Gamma^M \partial_M F(\phi) \gamma^5 \Psi$ introduced in Ref. \cite{XuZengGuang}.

This paper is organized as follows. In next section, we review the scalar-tensor brane model proposed in \cite{YangKe}, including the two solutions of the brane system and
the mass spectrum of gravity on the brane.
In Sec.~\ref{STbrane_SecBulkMatterFieldLocalize}, localization
and mass spectra of the scalar, vector, KR, and fermion fields are investigated for the two solutions.
Finally, we summarize our results in
Sec.~\ref{STbrane_secConclusion}.

\section{Review of the {scalar-tensor} brane model}
\label{STbrane_SecModel}

Let us consider a scalar-tensor brane generated by a real
scalar field $\phi$ nonminimally coupled to gravity.
The action for such a system is given by \cite{YangKe}
\begin{equation}
S = \frac{M_*^3}{2} \int d^5 x  \sqrt{-g} e^{k \phi} \left[ R
 -\left( 3 + 4k \right) \left( \partial \phi \right)^2 \right],
\label{STbrane_action}
\end{equation}
where $R$ is the
five-dimensional scalar curvature and $M_*$ is the fundamental scale of gravity. The line-element for a
five-dimensional space-time describing a Minkowski brane with an $S^1/Z_2 $ orbifold extra dimension is assumed
as
\begin{eqnarray}
ds^2 = a^2(z)\left( \eta_{\mu\nu}dx^\mu dx^\nu+ dz^2 \right)
 =e^{2A(z)}\left( \eta_{\mu\nu}dx^\mu dx^\nu+ dz^2 \right),  \label{STbrane_linee}
\end{eqnarray}
where the conformal coordinate $z \in [-z_b, z_b] $ is related to the physical extra dimension coordinate $y$ by a coordinate transformation $dy = a(z)dz$.
According to the symmetric of the space-time, the solution of the background scalar field $\phi$ depends on extra dimension only. The field equations in the bulk
derived from the action (\ref{STbrane_action}) under these assumptions
reduce to the following second-order differential equations:
\begin{eqnarray}
   2k \phi{''}
   + \left( 2k^2+ 4k+ 3\right)\phi'^2
   + 4k (\ln{a})'\phi'+ 6 \frac{a''}{a}
  & = & 0 ,       \label{STbrane_CoupleField_Eq1} \\
   \left[\phi'-2(\ln{a})'\right] \left[( 4k+ 3) \phi'+ 6(\ln{a})' \right]
  & = & 0,         \label{STbrane_CoupleField_Eq2} \\
   ( 4k+ 3)\left[2\phi{''}
                 + k \phi'^2
                 + 6(\ln{a})'\phi'
           \right]
     - 4k\left[2\frac{a''}{a}+ (\ln{a})'^2\right]
  & = & 0,           \label{STbrane_CoupleField_Eq3}
\end{eqnarray}
where the prime denotes the derivative with respect to the coordinate $z$. From Eq. (\ref{STbrane_CoupleField_Eq2}), one has
\begin{eqnarray}
 \phi'=2(\ln{a})' , ~~ \text{or}~~
 ({4k+3})\phi'=-{6}(\ln{a})'. \label{STbrane_Eq4TwoCases}
\end{eqnarray}
We do not consider the trivial case $k = -3/4$ for the second equation in (\ref{STbrane_Eq4TwoCases}) because the corresponding solution for $a(z)$ is just a constant.
The following two independent brane solutions were found in Ref. \cite{YangKe}.

Solution I:
\begin{subequations} \label{STbrane_STbraneSolution1}
\begin{eqnarray}
 a \left( z \right)
    & = & \left( 1+ \beta |z| \right)^\frac{1}{2k+3}, \\
 \phi \left( z \right)
    & = & \frac{2}{2k+3} \ln \left( 1+ \beta |z| \right),
\end{eqnarray}
\end{subequations}
where the parameters satisfy $\beta > 0$ and $k < -3/2$.

Solution II:
\begin{subequations}\label{STbrane_STbraneSolution2}
\begin{eqnarray}
 a \left( z \right)
    & = & \left( 1+ \beta |z| \right)^\frac{4k+3}{3(2k+3)}, \\
 \phi \left( z \right)
    & = & -\frac{2}{2k+3} \ln \left( 1+ \beta |z| \right).
\end{eqnarray}
\end{subequations}
where $\beta > 0$ and $-3/2 < k < -3/4$.

In the model, there are two thin branes: a positive tension brane located at the origin $z = 0$ and a negative one at another orbifold fixed poiont $z=z_b$, which is the same as the case of the RS1 model. However, our world in this model is located at the positive tension brane rather than on the negative one in order to solve the gauge hierarchy problem. As a result, a correct Friedmann-like equation on the brane can be obtained (for detail see Ref.~\cite{YangKe}).

Stability and the zero mode of the gravitational perturbation on the scalar-tensor brane have been analyzed in Ref. \cite{YangKe}.
Here we give a brief review of the mass spectrum of the gravitational perturbation for the
two solutions (solution I and solution II).

The analyzing of a full set of fluctuations
of the metric around the background is very complex. However, the problem
can be simplified when one only considers the transverse and traceless (TT) part of the metric fluctuation. So,
we consider the following TT tensor perturbation of the metric (\ref{STbrane_linee}):
\begin{eqnarray}
 ds^2 =  a^2(z)\left[ \left( \eta_{\mu \nu}
      + \bar{h}_{\mu \nu}(x,z) \right)dx^\mu dx^\nu
      + dz^2 \right],
\label{STbrane_pertubation_metric}
\end{eqnarray}
where the tensor perturbation $\bar{h}_{\mu \nu}$
satisfies the TT condition \cite{dewolfe}:
${\bar{h}_\mu}^\mu=\partial^\nu \bar{h}_{\mu \nu}=0$.
The equation for $\bar{h}_{\mu\nu}$ is given by \cite{YangKe}
\begin{equation}
  \bar{h}''_{\mu \nu}+3 \frac{a'}{a} \bar{h}'_{\mu \nu}
      + k\phi' \bar{h}'_{\mu \nu}+ \Box^{(4)}\bar{h}'_{\mu \nu}=0.
      \label{STbrane_Eqs_of_bar_h_mu_nu}
\end{equation}
By performing the following decomposition
\begin{equation}
  \bar{h}_{\mu \nu}(x,z) = \varepsilon_{\mu\nu}(x) {J}^{-\frac{3}{2}}(z) h(z),
\label{STbrane_decompsoseh_mu_nu}
\end{equation}
where the function ${J}(z)$ is defined as ${J(z)}=a(z)e^{k\phi/3}$, and its explicit expressions
for solution I and solution II are same:
\begin{equation}
      {J}(z) = \left( 1+ \beta |z| \right)^\frac{1}{3},
\label{STbrane_KK_G_warpe_factor}
\end{equation}
we get from Eq. (\ref{STbrane_Eqs_of_bar_h_mu_nu}) the Klein-Gordon equation $\Box^{(4)}\varepsilon_{\mu\nu}(x)=m^2\varepsilon_{\mu\nu}(x)$ for the four-dimensional gravity $\varepsilon_{\mu\nu}(x)$, and
a Schr\"{o}dinger-like equation for the KK mode $h(z)$:
\begin{equation}
      \left(-\partial_z^2 + V_{\text{g}}(z)\right) h(z) = m^2 h(z).
\label{STbrane_KK_G_Schdinger_Eqs}
\end{equation}
Here $m$ is the mass of the four-dimensional graviton (the gravitational KK excitation) and the effective potential is given by
\begin{eqnarray}
 V_{\text{g}}(z)=\frac{3}{2} \frac{{J}''}{{J}}+ \frac{3}{4} \frac{{J}'^2}{{J}^2}.\label{STbrane_KK_G_potential}
\end{eqnarray}
The explicit expressions of $V_{\text{g}}(z)$ for both solutions are the same:
\begin{eqnarray}
 V_{\text{g}}(z)=-\frac{{\beta^2}}{4(1+ \beta |z|)^2}
       + \beta \delta(z)-\frac{\beta}{1+\beta z_b}\delta(z-z_b),\label{STbrane_G_potential_explicit}
\end{eqnarray}
where the second delta function in the right hand side of the above equation comes from the $Z_2$ symmetry with respect to the brane located at $z=z_b$.
Then, the values of the effective potential $V_{\text{g}}(z)$ at $z=0$ and $z =z_b$ are
\begin{eqnarray}
 V_{\text{g}}(0) &=& -\frac{\beta^2}{4}+ \beta \delta(0) >0,\\
 V_{\text{g}}( z_b) &=&-\frac{\beta^2}{4(1+ \beta z_b)^2} - \frac{\beta}{1+\beta z_b} {\delta(0)}<0,
\end{eqnarray}
from which we can deduce that the gravity zero mode may be localized on the negative tension brane located at $z=z_b$.

By setting $m=0$ in Eq. (\ref{STbrane_KK_G_Schdinger_Eqs}), one can easily get the normalized zero mode
\begin{eqnarray}
  h_{0}(z) =C_0 {J^{\frac{3}{2}}}=C_0 a^{\frac{3}{2}} e^{k\phi/2}= \sqrt{\frac{1+ \beta |z|}{2z_b+ \beta z_b^2}}.
\end{eqnarray}
It is clear that the zero mode is localized on the negative tension brane for the finite $z_b$, but cannot be normalized anymore when the extra dimension is infinity,
which is very different from the RS1 model.

By solving the Schr\"{o}dinger-like equation (\ref{STbrane_KK_G_Schdinger_Eqs}),
the spectrum of gravity is \cite{YangKe}:
\begin{eqnarray}
  m_{n}(z) = \frac{x_{n}}{z_b + \frac{1}{\beta}},
\end{eqnarray}
where $x_n$ satisfies $J_1{(x_n)}=0$, and $x_1$=3.83, $x_2$=7.02,
$x_3$=10.17, $\cdots$. Here, $J_1{(x)}$ is the Bessel function of the first kind.

In order to solve the hierarchy problem, we need to set all the fundamental parameters including the five-dimensional scale of gravity $M_*$, the parameter $\beta$, and the Higgs vacuum expectation value $v_0$, to be about the TeV scale and $\beta z_b \approx 10^{16}$ \cite{YangKe}. So we have $\beta\approx 10^{12}$eV, $z_b\approx10^{4}$eV$^{-1}$, and $m_{n}(z) = {x_{n}} \times 10^{-4}$ eV. Note that the mass spectra for both the two brane solutions are the same and the mass spacing of KK gravitons is much smaller than that of the RS1 model with the TeV scale spacing. So we cannot distinguish the two brane solutions from the gravitational mass spectrum. However, we will see in the following sections that the mass spectra of various matter fields are different for the two brane solutions, and so they are not physically equivalent.

\section{Localization and mass spectra of various matters on the scalar-tensor brane}
\label{STbrane_SecBulkMatterFieldLocalize}

In this section, we would like to investigate localization
and mass spectra of various bulk matter fiedls on the scalar-tensor {brane}
by deriving the effective potentials of the KK modes of various bulk matter fiedls in the corresponding
Schr\"{o}dinger-like equations.
The bulk matter fiedls include the spin-0 scalar, spin-1 vector, KR, and spin-1/2 fermion fields. For the spin-1/2 fermion field, we will introduce a new scalar-fermion coupling in order to localize the fermion on the {brane}.
Here, we implicitly assume that the various bulk matter fields
considered below are perturbations around the background space-time and they
make little backreaction to the bulk
energy so that the brane solutions given in section \ref{STbrane_SecModel}
remain valid.

\subsection{Spin-0 scalar field}
\label{STbrane_sec3.1}

We first consider the localization of the massless real scalar field on the scalar-tensor  brane reviewed in
section \ref{STbrane_SecModel}.
The five-dimensional action for a massless real scalar field coupled to the dilaton field $\phi$ is given by
\begin{equation}
S_{\text{s}} = -\frac{1}{2}\int d^5 x  \sqrt{-g}~ e^{\lambda\phi} g^{MN} \partial_M \Phi \partial_N \Phi,
\label{STbrane_scalar_action}
\end{equation}
where $\lambda$
is the coupling constant between the scalar and dilaton fields.
By considering the action (\ref{STbrane_scalar_action}) and the metric (\ref{STbrane_linee}), the equation of motion for the scalar field is read as
\begin{eqnarray}
 \partial_{P} \left(\sqrt {-g} e^{\lambda\phi} \partial^{P} \Phi \right)=0.
\end{eqnarray}

{
We make the KK decomposition of the scalar field
\begin{eqnarray}
 \Phi(x^{\mu},z)=\sum_n \phi_n(x^\mu) \bar{\chi}_n(z) =\sum_n \phi_n(x^\mu) \chi_n(z) e^{-H_i A(z)}, ~~~~(i=I,II)
\end{eqnarray}
where  $H_I = \frac{2\lambda+3}{2}$ and $H_{II}= \frac{3(4k+3)-6\lambda}{2(4k+3)}$ for solutions I  and II, respectively. Note that the function $e^{-H_i A(z)}$ in the above decomposition is to eliminate the first-order term in the equation of motion of the redefined function $\chi_n(z)$ (see Ref. \cite{Liu0708} for the detail calculation). Then, the equation of motion for the extra-dimensional part $\chi_n (z)$ of the scalar KK mode is recast into the following Schr\"{o}dinger-like equation
\begin{equation}
\left[ -\partial_z^2 + V_{\text{s}}^i(z) \right] \chi_n(z)= m_n^2 \chi_n(z), \label{STbrane_scalar_equation}
\end{equation}
where $m_n$ is the mass of the scalar KK mode $\phi_n(x^\mu)$ and the effective potential $V_{\text{s}}^i(z)$ is given by
\begin{eqnarray}
 V_{\text{s}}^i(z)=H_i\partial_{z}^2 A(z)
        +[H_i\partial_{z}A(z)]^2.  ~~~~(i=I,II)
\label{STbrane_KK_scalar_potential}
\end{eqnarray}
By introducing the orthonormality condition
\begin{eqnarray}
\int_{-z_{b}}^{+z_{b}} \chi_m(z) \chi_n(z)dz=\delta_{mn}, \label{STbrane_scalar_orthonormality_condition}
\end{eqnarray}
we get the four-dimensional effective action of a massless and a series of massive scalar fields from the five-dimensional one:
\begin{eqnarray}
 S_{\text{s}} = -\frac{1}{2}\sum_n \int d^4 x ( \eta^{\mu\nu} \partial_\mu \phi_n \partial_\nu \phi_n + m_n^2 \phi_{n}^2 ).
 \label{STbrane_scalar_4D_action}
\end{eqnarray}

Substituting the explicit forms of $H_i$ and  $A(z)$ into
Eq. (\ref{STbrane_KK_scalar_potential}), we get the explicit expression of the effective potential $V_{\text{s}}^i(z)$:
\begin{eqnarray}
 V_{\text{s}}^i(z)&=&\frac{(2\lambda + 3 )[2\lambda - (4k + 3)]\beta^2}{4 ( 2k+3 )^2 (1+ \beta |z|)^2}
        + \frac{c_i \beta }{2k+3}
          \Big[\delta(z)-\frac{1}{1+\beta z_b}\delta(z-z_b)\Big],
           \label{STbrane_scalar_potential_V0}
\end{eqnarray}
where $c_{I}=2\lambda + 3$ and $c_{II}=-2\lambda + 4k +3$.
The values of $V_{\text{s}}^i(z)$ at $z = 0 $ and $|z|= z_{b} $ read
\begin{eqnarray}
V_{\text{s}}^i(0)&=&\frac{(2\lambda + 3 )[2\lambda - (4k + 3)]\beta^2}{4 ( 2k+3 )^2}
        +\frac{c_i\beta }{2k+3}\delta(0), \\
V_{\text{s}}^i(z_{b})&=&\frac{(2\lambda + 3 )[2\lambda - (4k + 3)]\beta^2}
        {4 ( 2k+3 )^2 (1+ \beta z_b)^2}
       - \frac{c_i\beta }{(2k+3)(1+\beta z_b)}\delta(0).
\end{eqnarray}

In order to localize the scalar zero mode on the positive tension brane located at $z=0$, the effective potential $V_{\text{s}}^i(z)$ should be negative at $z=0$.
The condition is turned out to be
\begin{eqnarray}
 \lambda > \left\{ \begin{array}{ll}
                     -{3}/{2} & \text{~~~for solution I}, \\
                     ({4k+3})/{2} &  \text{~~~for solution II}.
                   \end{array}
           \right.
    \label{STbrane_scalar_localized_condition}
\end{eqnarray}

By setting $m=0$ in Eq. (\ref{STbrane_scalar_equation}) and noting that the boundaries of the extra dimension are at $z=0$ and $z=z_b$,
we get the normalized zero mode of the scalar field:
\begin{eqnarray}
 \chi_{0}(z)
             =\left\{ \begin{array}{lll}
                     \sqrt {\frac{\beta (\lambda+k+3)}
                           {(2k+3)\big[(1+\beta z_b)^\frac{2(\lambda+k+3)}{2k+3}-1\big]}}
                     \left( 1+ \beta |z| \right)^\frac{2\lambda + 3}{2(2k+3)}, & \text{~for solution I}~~ (\lambda\neq -k-3), \\
                     \sqrt{\frac{\beta(-\lambda+3k+3)}
                                {(2k+3)  \big[(1+\beta z_b)
                                  ^\frac{2(-\lambda+3k+3)}
                                 {2k+3}-1\big]}}
                        ( 1+ \beta |z| )^\frac{-2\lambda +  4k+3}{2(2k+3)}, &  \text{~for solution II}~(\lambda\neq 3k+3).
                   \end{array}
           \right.
             \label{STbrane_scalar_zero_mode}
\end{eqnarray}
It can be checked that under the condition (\ref{STbrane_scalar_localized_condition}), the scalar zero mode has a maximum at $z=0$, and so the scalar zero mode can be localized on the positive tension brane.
We note here that, when there is no coupling between the scalar and dilaton fields ($\lambda=0$),
the scalar zero mode is also localized on the positive tension brane.
Besides, if the extra dimension is infinite,
{the normalization condition
$\int_{-\infty}^{+\infty} \chi_{0}^2(z) dz =1$ should be satisfied,
which shows that the localization condition for the scalar zero mode is much stronger: $\lambda>-k-3$} for solution I and $\lambda> 3k+3$ for solution II.

When the coupling constant $\lambda= -k-3$ for solution I and $\lambda= 3k+3$ for solution II, the normalized scalar zero mode $\chi_{0}(z)$ is
\begin{eqnarray}
   \chi_{0}(z)=\sqrt{\frac{\beta}{2 \ln(1+\beta z_b)}}\frac{1}{\sqrt{1+\beta |z|}}.
 \label{STbrane_scalar_special_zero_mode}
\end{eqnarray}
For this special coupling, the zero mode can also be localized on the positive tension brane
when $z_b$ is finite.
But it {cannot anymore}  when the extra dimension is infinite.

In order to get the massive spectrum of the scalar KK modes, we assume that the extra dimension is finite.
The $Z_2$ symmetry requires that the KK modes satisfy $\partial_z(e^{-H_i A(z)} \chi_n)=0$
at the boundary $z=0$, with which the general solution of Eq. (\ref{STbrane_scalar_equation}) is turned out to be
\begin{eqnarray}
 \chi_n(z)=N(1+\beta |z|)^\frac{1}{2}
     \Big[ \text{J}_{P_{\text{s}}}(\bar{z})
          +\alpha_{\text{s}}^i \text {Y}_{P_{\text{s}}}(\bar{z})
     \Big],
 \label{STbrane_scalar_massive_Solution}
\end{eqnarray}
where $N$ is the normalization coefficient,
$\text{J}_{P_{\text{s}}}(z)$ and $\text{Y}_{P_{\text{s}}}(z)$ are respectively
the Bessel functions of first and second kinds, and
\begin{eqnarray}
  \bar{z}&\equiv& m_n(|z|+\frac{1}{\beta}),~
  {P_{\text{s}}}\equiv {\frac{\sqrt{1-Q_{\text{s}}}}{2}},~
   Q_{\text{s}} \equiv \frac{(2\lambda+3)(-2\lambda+4k+3)}{(2k+3)^2},~\\
  \alpha_{\text{s}}^I &\equiv&
           -\frac{\text{J}_{P_{\text{s}}-1}(\frac{m_n}{\beta})}
                 {\text{Y}_{P_{\text{s}}-1}(\frac{m_n}{\beta})}.\\
  \alpha_{\text{s}}^{II} &\equiv& -\frac{\big[(2k+3)m_n +2\beta(\lambda-k)\big]
       \text{J}_{P_{\text{s}}-1}(\frac{m_n}{\beta}) -(2k+3)m_n \text{J}_{P_{\text{s}}+1} (\frac{m_n}{\beta})}
             {\big[(2k+3)m_n +2\beta(\lambda-k)\big] \text{Y}_{P_{\text{s}}-1}(\frac{m_n}{\beta}) -(2k+3)m_n \text{Y}_{P_{\text{s}}+1} (\frac{m_n}{\beta})}.
\end{eqnarray}

The $Z_2$ symmetry also requires that the KK modes satisfy $\partial_z(e^{-H_{i}A(z)} \chi_n(z))=0$
at another boundary $z=z_b$. When we consider the light modes in the long range case, i.e.,
$m_n/\beta \ll 1$ and $1+\beta z_b\gg 1$, the spectrum of the
scalar KK modes can be determined by the following equation
\begin{eqnarray}
     \frac{\text{J}_{P_{\text{s}}} (\bar{z_b})
              +\alpha_{\text{s}}^I \text{Y}_{P_{\text{s}}}(\bar{z_b})
          }
          {\text{J}_{P_{\text{s}}+1} (\bar{z_b})
               +\alpha_{\text{s}}^I \text{Y}_{P_{\text{s}}+1} (\bar{z_b})
          }
     &=&  \frac{(2k+3) \bar{z_b}}
        {k+3+(P_{\text{s}}-1)(2k+3)-\lambda} ~~\text{~for solution I},\\
    \frac{\text{J}_{P_{\text{s}}} (\bar{z_b})
            +\alpha_{\text{s}}^{II} \text{Y}_{P_{\text{s}}} (\bar{z_b}) }
           {\text{J}_{P_{\text{s}}+1} (\bar{z_b})
            +\alpha_{\text{s}}^{II} \text{Y}_{P_{\text{s}}+1} (\bar{z_b})  }
    &=&  \frac{(2k+3) \bar{z_b}}{k+3+(P_{\text{s}}-1)(2k+3)+\lambda } ~~\text{~for solution II},
\end{eqnarray}
where $\bar{z_b}\equiv m_n(z_b+{1}/{\beta})$.

We can obtain the mass spectrum of the scalar field by numerical calculation.
For example, for solution I, when the parameters are set to $\beta=10^{12}\text{ev},~z_b=10^4\text{ev}^{-1},~\lambda=1,~k=-2$,
the mass spectrum is $m_1=0.51\times10^{-3} \text{ev},~m_2=0.84\times10^{-3} \text{ev},
~m_3=1.16\times10^{-3} \text{ev},~m_4=1.48\times10^{-3} \text{ev}$,~$\cdots$.
The explicit spectrum of the scalar field is shown in figs.~\ref{STbrane_fig_Spectrum_Scalar_SolutionI} and \ref{STbrane_fig_Spectrum_Scalar_SolutionII} for different $\lambda$ and $k$.
For solution I, the mass gap between the massless mode and the first massive KK mode increases with
the coupling constant $\lambda$ and the non-minimal coupling parameter $k$.
Besides, the mass spectrum is relatively sparse at
the lower excited states but approaches equidistant for the higher excited states.
For solution II, the mass gap between the zero mode and the first massive KK mode is increased with the coupling constant $\lambda$ but decreased with the parameter $k$, which is different from that in solution I.

A natural question is that are there particular values of parameters in which both solutions I and II lead to the same mass spectra?
The answer is yes. It is not difficult to see that, when $\lambda=k$, the effective potentials for both solutions are the same:
\begin{eqnarray}
 V_{\text{s}}^I(z)=V_{\text{s}}^{II}(z)=-\frac{{\beta^2}}{4(1+ \beta |z|)^2}+
         \beta \Big[\delta(z)-\frac{\delta(z-z_b)}{1+\beta z_b}\Big],
\end{eqnarray}
and the boundary conditions
\begin{eqnarray}
 \partial_z(e^{-H_i A(z)} \chi_n)|_{z=0,z_b}=\partial_z(1+\beta |z|)^\frac{1}{2}|_{z=0,z_b}=0
\end{eqnarray}
are also the same. However, the scalar zero mode for this special case is localized on the negative tension brane but not the positive one for both solutions, because letting $\lambda=k$ will contradict with the localization condition (\ref{STbrane_scalar_localized_condition}) on the positive tension brane.}

\begin{figure}[htb]
\subfigure[$k=-2$]{
\includegraphics[width=7cm,height=4.5cm]{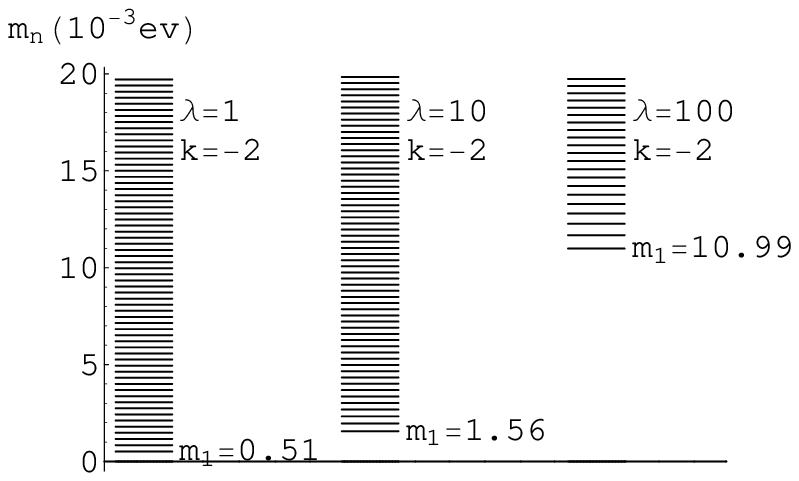}}
\subfigure[$\lambda=10$]{
\includegraphics[width=7cm,height=4.5cm]{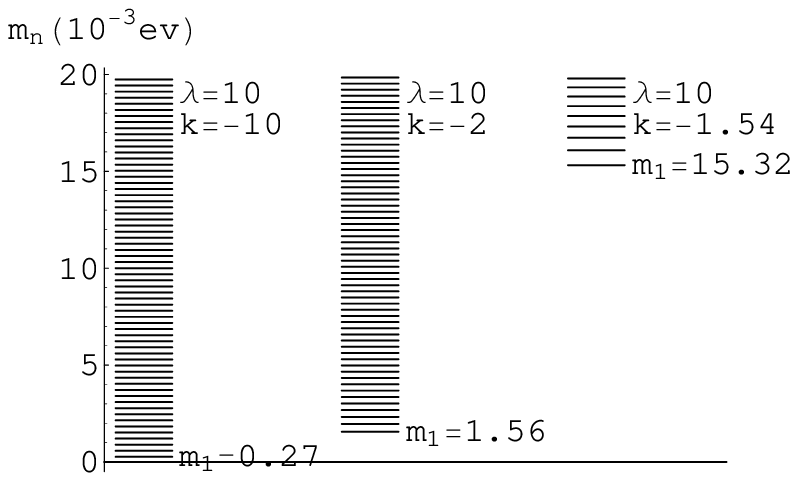}}
 \caption{The mass spectrum of the scalar KK modes for solution I.
 The parameters  are set to $\beta=10^{12}$\text{ev} and $z_b=10^4$\text{ev}$^{-1}$.}
\label{STbrane_fig_Spectrum_Scalar_SolutionI}
\end{figure}

\begin{figure}[htb]
\subfigure[$k=-1$]{
\includegraphics[width=7cm,height=4.5cm]{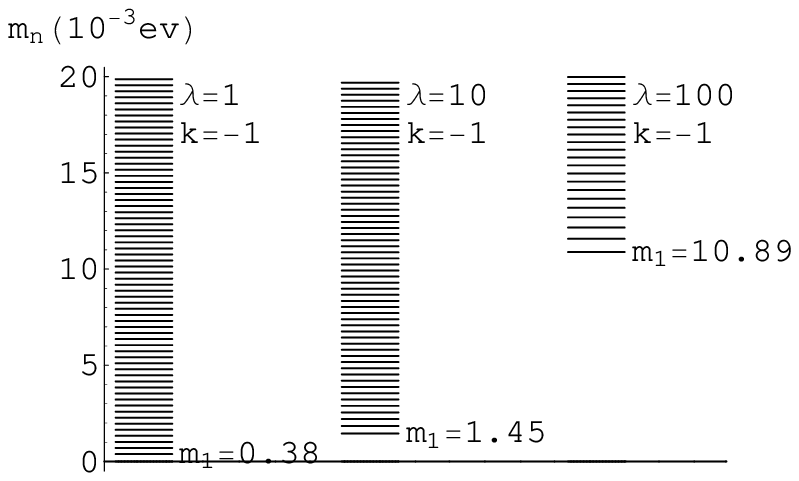}}
\subfigure[$\lambda=10$]{
\includegraphics[width=7cm,height=4.5cm]{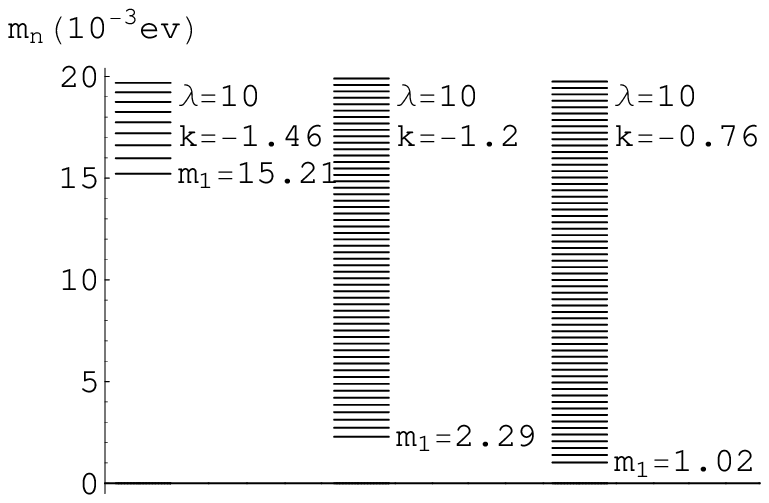}}
 \caption{The mass spectrum of the scalar KK modes for solution II.
 The parameters  are set to $\beta=10^{12}$\text{ev} and $z_b=10^4$\text{ev}$^{-1}$.}
\label{STbrane_fig_Spectrum_Scalar_SolutionII}
\end{figure}

\subsection{Spin-1 vector field}
\label{STbrane_secFermionZeroMode}

{Now we turn to the spin-1 vector field.} Let's begin with the five-dimensional action of a
vector field  {coupled to} a dilaton field $\phi$:
\begin{equation}
S_{\text{v}} = -\frac{1}{4} \int d^5 x  \sqrt{-g} e^{\tau\phi}g^{MN} g^{RS}F_{MR}F_{NS},
\label{STbrane_vector_action1}
\end{equation}
where $F_{MN}=\partial_{M}A_{N}-\partial_{N}A_{M}$ is the field strength tensor and $\tau$
is the coupling constant between the vector and dilaton fields.
 {Considering the explicit form of} the metric (\ref{STbrane_linee}), the equations of motion  {for the vector field are read as}
\begin{eqnarray}
  e^{\tau\phi}\partial_{\nu} \left(  \eta^{\nu\rho} \eta^{\mu\lambda} F_{\rho\lambda}\right) +\eta^{\mu\lambda} a^{-1}(z)\partial_4 \left(a(z)e^{\tau\phi}F_{4\lambda}\right) &=& 0,   \nonumber \\
  e^{\tau\phi} \partial_\mu \left( \eta^{\mu\nu} F_{\nu4} \right)  &=& 0.
\end{eqnarray}
By using gauge freedom and the $Z_2$ symmetry of extra dimension, we can set the fourth component $A_4=0$.
Next, we investigate localization of the zero mode and  KK mass spectrum of the vector  filed
on the scalar-tensor  brane for the two brane solutions.

{
By performing the following KK decomposition for the vector field
\begin{eqnarray}
 A^{\lambda}(x^\mu,z)=\sum_n a_n^{\lambda}(x^\mu) \rho_n(z)e^{-U_i A(z)},~~~~(i=I,II)
\end{eqnarray}
where $U_I=\frac{2\tau+1}{2}$ for solution I and $U_{II}= \frac{-6\tau+4k+3}{2(4k+3)}$ for
solution II.
One can show that the extra dimension part $\rho_n (z)$ of the vector KK mode satisfies the following Schr\"{o}dinger-like equation
\begin{equation}
\left[ -\partial_z^2 + V_{\text{v}}^i(z) \right] \rho_n(z)= m_n^2 \rho_n(z), \label{STbrane_vector_equation}
\end{equation}
where the effective potential $V_{\text{v}}^i(z)$ is given by
\begin{eqnarray}
 V_{\text{v}}^i(z)=(U_i \partial_{z}A)^2
        +U_i \partial_{z}^2 A.
 \label{STbrane_KK_vector_potential}
\end{eqnarray}
With the orthonormality condition
\begin{eqnarray}
\int_{-z_{b}}^{+z_{b}} \rho_m(z) \rho_n(z)dz=\delta_{mn}, \label{STbrane_vector_orthonormality_condition}
\end{eqnarray}
we can get the four-dimensional effective action for {a series of} vector fields:
 \begin{eqnarray}
 S_{\text{v}} = \sum_{n}\int d^4 x \left(-\frac{1}{4}\eta^{\mu\alpha}\eta^{\nu\beta}f^{(n)}_{\mu\nu}
 f^{(n)}_{\alpha\beta} - \frac{1}{2}m^2_{n} \eta^{\mu\nu} a^{(n)}_{\mu} a^{(n)}_{\nu} \right),
 \label{STbrane_vector_4D_action}
\end{eqnarray}
where $f^{(n)}_{\mu\nu}=\partial_\mu a^{(n)}_\nu - \partial_\nu a^{(n)}_\mu$ is the four-dimensional vector field strength tensor.

The explicit expression of the effective potential $V_{\text{v}}^i(z)$ read
\begin{eqnarray}
 V_{\text{v}}^{I}(z)&=&\frac{(2\tau+1)\left( 2\tau - 4k- 5  \right) \beta^2}{4 \left( 2k+3 \right)^2 ( 1+ \beta |z|)^2}
        +\frac{(2\tau+1)\beta }{2k+3}\Big[ \delta(z)- \frac{\delta(z-z_b)}{1+\beta z_b} \Big].
 \label{STbrane_vector_potential_V1_1}  \\
  V_{\text{v}}^{II}(z)&=&\frac{ (6\tau + 8k+15)(6\tau-4k-3)\beta^2}
                   {36 (2k+3)^2 (1 + \beta |z|)^2}
        + \frac{\beta(- 6\tau+ 4k+3)}{3( 2k+3)}\Big[ \delta(z)- \frac{\delta(z-z_b)}{1+\beta z_b}  \Big].
 \label{STbrane_vector_potential_V1_2}
\end{eqnarray}
The values of $V_{\text{v}}^i(z)$ at $z = 0 $ are
\begin{eqnarray}
V_{\text{v}}^I(0)&=&\frac{(2\tau+1)\left( 2\tau - 4k- 5  \right) \beta^2}{4 \left( 2k+3 \right)^2 }
        +\frac{(2\tau+1)\beta }{2k+3} \delta(0), \\
V_{\text{v}}^{II}(0)&=&\frac{ (6\tau + 8k+15)(6\tau-4k-3)\beta^2}{36 (2k+3)^2 }
        + \frac{\beta(- 6\tau + 4k+3 )}{3( 2k+3)}\delta(0).
\end{eqnarray}
In order to  localize the vector zero mode on the positive tension brane,
the effective potential $V_{\text{v}}^i(z)$
should be  negative or have a well-like shape near $z=0$.
The condition is turned out to be
\begin{eqnarray}
 \tau > \left\{ \begin{array}{ll}
                     -{1}/{2} & \text{~~~for solution I}, \\
                     ({4k+3})/{6} &  \text{~~~for solution II}.
                   \end{array}
           \right.
    \label{vector_norcondition1_1}
\end{eqnarray}
By setting $m=0$ in Eq. (\ref{STbrane_vector_equation}),
we get the normalized vector zero mode:
\begin{eqnarray}
 \rho_{0}(z) =\left\{ \begin{array}{lll}
                     \sqrt{\frac{\beta(\tau+k+2)}{(2k+3)
                         \big[(1+\beta z_b)^\frac{2(\tau+k+2)}{2k+3}-1\big]}}
                         \left( 1+ \beta |z|\right)^\frac{2\tau+1}{2(2k+3)},
                     & \text{~for solution I}~~ (\tau \neq -k-2), \\
                    \sqrt{\frac{\beta(-3\tau+5k+6)}
                         {3(2k+3)\big[(1+\beta z_b)^\frac{2(-3\tau+5k+6)}{3(2k+3)}-1\big]}}
                        \left( 1 + \beta |z|\right)^{\frac{ - 6\tau+4k+3}{6(2k+3)}},
                    &  \text{~for solution II}~(\tau\neq\frac{5k+6}{3}).
                   \end{array}
           \right.             
\end{eqnarray}
It can be seen that, when the extra dimension is finite, the vector zero mode can be localized on the positive tension brane providing that the coupling constant between  the vector and dilaton fields satisfies the condition (\ref{vector_norcondition1_1}).
Clearly, the vector zero mode still can be localized on the brane even if there is no coupling
between the vector and dilaton fields.
When the extra dimension size is infinite,
the localization condition becomes much stronger, i.e., $\tau > -k-2$ for solution I and  $\tau > \frac{5k+6}{3}$ for solution II.

When $\tau= -k-2$ for solution I and $\tau= \frac{5k+6}{3}$ for solution II, the normalized vector zero mode $\rho_{0}(z)$ is
\begin{eqnarray}
 \rho_{0}(z)=\sqrt{\frac{\beta}{2 \ln(1+\beta z_b)}}\frac{1}{\sqrt{1+\beta |z|}}.
\label{STbrane_special_zeromode_vector}
\end{eqnarray}
It is localized on the positive tension brane only for the case of finite extra dimension.

Next, we will investigate the massive KK modes of the vector field by assuming that the extra dimension
is compact and finite. With the boundary condition $\partial_z(e^{-U_i A(z)} \rho_n)|_{z=0}=0$,
 we get the general solution of Eq. (\ref{STbrane_vector_equation}):
\begin{eqnarray}
 \rho_n(z)={N} (1+\beta |z|)^\frac{1}{2}
            \Big[
                 \text{J}_{P_{\text{v}}^i} (\bar{z}) +
            \alpha_{\text{v}}^i ~ \text {Y}_{P_{\text{v}}^i}(\bar{z})
            \Big],
\end{eqnarray}
where
\begin{eqnarray}
   \alpha_{\text{v}}^i&\equiv& -\frac{\text{J}_{P_{\text{v}}^i-1}(\frac{m_n}{\beta})}{\text{Y}_{P_{\text{v}}^i-1}(\frac{m_n}{\beta})},~~ {P_{\text{v}}^i} \equiv {\frac{\sqrt{1-Q_{\text{v}}^i}}{2}},\nonumber \\
   Q_{\text{v}}^I &\equiv& \frac{(-2\tau+4k+5)(2\tau+1)}{(2k+3)^2},~~
   Q_{\text{v}}^{II} \equiv \frac{(-6\tau+4k+3)(6\tau+8k+15)}{9(2k+3)^2}.
\end{eqnarray}
With another boundary condition at $z=z_b$: $\partial_z(e^{-U_i A(z)} \rho_n)|_{z=z_b}=0$,
the spectrum of the vector KK modes for light modes in long range case
is determined by
\begin{eqnarray}
  \frac{\text{J}_{P_{\text{v}}^I} (\bar{z_b}) +\alpha_{\text{v}}^I \text{Y}_{P_{\text{v}}^I} (\bar{z_b})}
      {\text{J}_{P_{\text{v}}^I+1} (\bar{z_b})+
                        \alpha_{\text{v}}^I \text{Y}_{P_{\text{v}}^I+1} (\bar{z_b})  }
    &=&\frac{(2k+3) \bar{z_b}}{3k+4+(P_{\text{v}}^I-1)(2k+3)-\tau }   ~~~~~~\text {~for solution I},\\
  \frac{\text{J}_{P_{\text{v}}^{II}} (\bar{z_b}) +\alpha_{\text{v}}^{II} \text{Y}_{P_{\text{v}}^{II}} (\bar{z_b}) }
      {\text{J}_{P_{\text{v}}^{II}+1} (\bar{z_b})+
                        \alpha_{\text{v}}^{II} \text{Y}_{P_{\text{v}}^{II}+1} (\bar{z_b})  }
    &=&\frac{3(2k+3) \bar{z_b}}{7k+12+(P_{\text{v}}^{II}-1)(6k+9)+3\tau } ~~\text {~for solution II}.~~~~
\end{eqnarray}
The mass spectrum for the vector field is plotted in figs. \ref{STbrane_fig_Spectrum_Vector_SolutionI} and \ref{STbrane_fig_Spectrum_Vector_SolutionII}, which show that the mass gap between the massless and first massive modes increases with
the coupling constant $\tau$ and the parameter $k$ for solution I, but increases with the coupling constant $\tau$ and decreases with the parameter $k$ for solution II.
Besides, the spectrum interval approaches a constant for
higher excited states.

Similar to the scalar field, when $\tau$ takes a particular value i.e., $\tau=k/3$, we have
\begin{eqnarray}
 V_{\text{v}}^{I}(z)=V_{\text{v}}^{II}(z)
             =-\frac{5 \beta^2}{36(1+ \beta |z|)^2}+
         \frac{\beta}{3} \Big[\delta(z)-\frac{\delta(z-z_b)}{1+\beta z_b}\Big].
\end{eqnarray}
So the mass spectra of the vector KK modes for both solutions I and II are the same. However, the vector zero mode for this special case cannot be localized on the positive tension brane any more because the condition (\ref{vector_norcondition1_1}) is not satisfied. It is not difficult to check that the vector zero mode is localized on the negative tension brane.}

\begin{figure}[htb]
\subfigure[$k=-2$]{
\includegraphics[width=7cm,height=4.5cm]{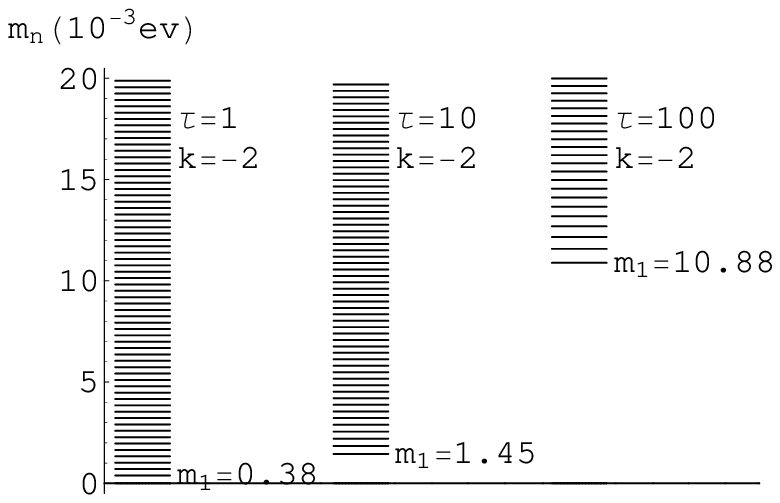}}
\subfigure[$\tau=10$]{
\includegraphics[width=7cm,height=4.5cm]{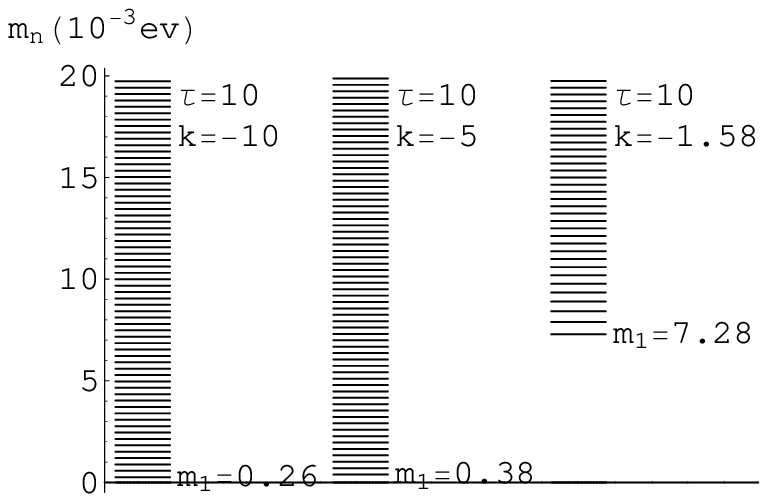}}
 \caption{The mass spectrum of the vector KK modes for solution I.
 The parameters  are set to $\beta=10^{12}$\text{ev} and $z_b=10^4$\text{ev}$^{-1}$.}
 \label{STbrane_fig_Spectrum_Vector_SolutionI}
\end{figure}
\begin{figure}[htb]
\subfigure[$k=-1$]{
\includegraphics[width=7cm,height=4.5cm]{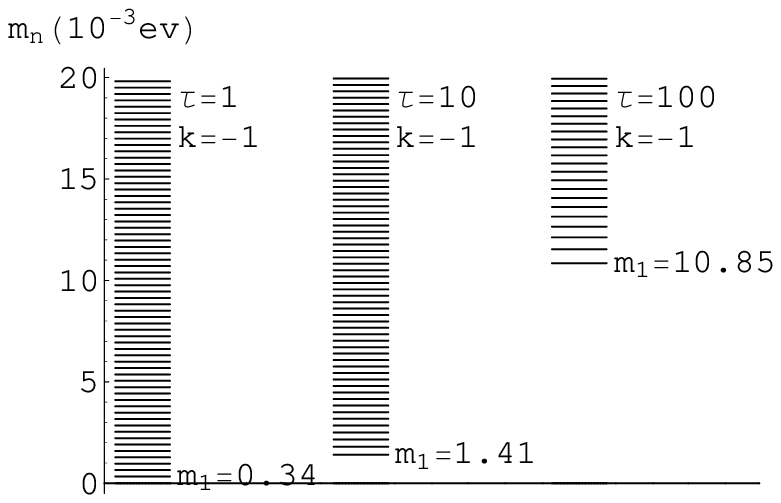}}
\subfigure[$\tau=10$]{
\includegraphics[width=7cm,height=4.5cm]{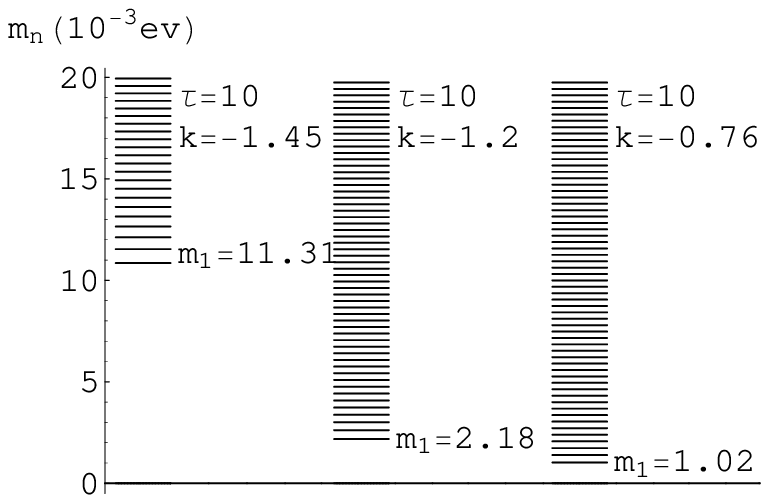}}
 \caption{The mass spectrum of the vector KK modes for solution II.
 The parameters  are set to $\beta=10^{12}$\text{ev} and $z_b=10^4$\text{ev}$^{-1}$.}
 \label{STbrane_fig_Spectrum_Vector_SolutionII}
\end{figure}

\subsection{Kalb-Ramond field}
\label{STbrane_secKRFields}

The action describing a five-dimensional KR field coupled with a dilaton field is given by
\begin{eqnarray}
 S_{\text{kr}}=-\int d^5x \sqrt{-g} e^{\zeta\phi} H_{MNL} H^{MNL},
 \label{STbrane_action_KR_fields}
\end{eqnarray}
where $H_{MNL}=\partial_{M}B_{NL}-\partial_{L}B_{NM}$ is the field strength of the KR field and $\zeta$ is the coupling constant between the KR
and dilaton fields. The equations of motion
are read as
\begin{eqnarray}
 e^{\zeta\phi} \partial_{\mu} \left(\sqrt{-g} H^{\mu\alpha\beta}\right) +
 \partial_4 \left( \sqrt{-g} e^{\zeta\phi} H^{4\alpha\beta}\right) & =& 0,\\
 e^{\zeta\phi} \partial_{\mu} \left(\sqrt{-g} H^{\mu 4 \beta}\right) & = &0.
 \label{STbrane_equation_KR_fields}
\end{eqnarray}
We choose the fourth component $B_{\alpha4} = 0$ by using gauge freedom.
Next, just like the vector field,
we will discuss localization and mass spectrum
of the KR KK modes for the two brane solutions considered in this paper.

With the KK decomposition of the KR field
\begin{eqnarray}
 B^{\alpha\beta}(x^{\mu},z)=\sum_n \hat b^{\alpha\beta}_{n}(x^{\mu})
              \Omega_n(z)e^{-R_i A(z)}, ~~~~(i=I,II)
\end{eqnarray}
where $R_{I}=\frac{2\zeta+7}{2}$ for solution I and $R_{II}=\frac{-6\zeta+7(4k+3)}{2(4k+3)}$ for solution II.
Providing the orthonormality condition for the KK modes $\Omega_m$ and $\Omega_n$:
\begin{eqnarray}
\int_{-z_{b}}^{+z_{b}} dz ~\Omega_m(z)\Omega_n(z)=\delta_{mn}, \label{STbrane_KR_orthonormality_condition}
\end{eqnarray}
we get the following Schr\"{o}dinger-like equation for the KR KK modes:
\begin{equation}
\left[ -\partial_z^2 + V_{\text{kr}}^i(z) \right] \Omega_n(z)= m_n^2 \Omega_n(z).
\label{STbrane_KR_equation}
\end{equation}
Here the effective potential $V_{\text{kr}}^i(z)$ is
\begin{eqnarray}
 V_{\text{kr}}^i(z)=[(R_i-4)\partial_{z}A]^2
        +(R_i-4) \partial_{z}^2 A.
\label{STbrane_KR_potential1}
\end{eqnarray}

Then the action of the KR field is reduced to
\begin{eqnarray}
 S_{\text{kr}}=-\sum_n \int d^4x \left( \eta^{\sigma' \sigma} \eta^{\alpha' \alpha} \eta^{\beta' \beta} \hat h^{(n)}_{\sigma' \alpha' \beta'} \hat h^{(n)}_{\sigma \alpha \beta}
 + \frac{1}{3} m_n^2 \eta^{\alpha' \alpha} \eta^{\beta' \beta} \hat b^{(n)}_{\alpha' \beta'} \hat b^{(n)}_{\alpha \beta} \right),
 \label{STbrane_KR_4D_action}
\end{eqnarray}
where $\hat h^{(n)}_{\sigma \alpha \beta} = \partial_\sigma \hat b_{\alpha \beta}- \partial_\beta \hat b_{\alpha \sigma}$ is
the field strength of the four-dimensional KR field.

The explicit expression of the effective potential $V_{\text{kr}}^i(z)$ for the KR KK modes is
\begin{eqnarray}
 V_{\text{kr}}^{I}(z)&=&
         \frac{(2\zeta - 1) \left( 2\zeta  - 4k- 7 \right) \beta^2}{4 \left( 2k+3 \right)^2 ( 1+ \beta |z|)^2}
        +\frac{(2\zeta - 1)\beta }{2k+3}\Big[ \delta(z)- \frac{\delta(z-z_b)}{1+\beta z_b}  \Big],\\~~~~~~~~
 V_{\text{kr}}^{II}(z)&=&
        \frac{(6\zeta  + 16k+ 21)(6\zeta  + 4k+ 3)\beta^2}{36 (2k+3)^2 (1 + \beta |z|)^2}
        - \frac{(6\zeta  + 4k+ 3)\beta}{3(2k+3)}\Big[ \delta(z)- \frac{\delta(z-z_b)}{1+\beta z_b} \Big].~~~~~~~~~
 \label{STbrane_KR_potential_V1_1}
\end{eqnarray}
The values of $V_{\text{kr}}^i(z)$ at $z = 0 $ are
\begin{eqnarray}
V_{\text{kr}}^I(0)&=&
        \frac{(2\zeta - 1) \left( 2\zeta  - 4k - 7\right) \beta^2}{4 \left( 2k+3 \right)^2 }
        +\frac{(2\zeta - 1)\beta }{2k+3} \delta(0), \\
V_{\text{kr}}^{II}(0)&=&
        \frac{(6\zeta  + 16k+ 21)(6\zeta  + 4k+ 3)\beta^2 }{36 (2k+3)^2 }
        - \frac{(6\zeta  + 4k+ 3)\beta}{3(2k+3)}\delta(0).
\end{eqnarray}
In order to get negative potential around $z = 0$, the coupling constant should satisfy the following constrain:
\begin{eqnarray}
 \zeta > \left\{ \begin{array}{ll}
                     {1}/{2} & \text{~~~for solution I}, \\
                     -({4k+3})/{6} &  \text{~~~for solution II}.
                   \end{array}
           \right.
    \label{STbrane_KR_norcondition1_11}
\end{eqnarray}
By setting $m=0$ in Eq. (\ref{STbrane_KR_equation}), we get the KR zero mode
\begin{eqnarray}
 \Omega_{0}(z) =\left\{ \begin{array}{lll}
                    \sqrt{\frac{\beta(\zeta+k+1)}
                  {(2k+3)\big((1+\beta z_b)^\frac{2(\zeta+k+1)}{2k+3}-1\big)}}
                     \left( 1+ \beta |z|\right)^\frac{2\zeta - 1}{2(2k+3)}
                     & \text{~for solution I}~~ (\zeta\neq-k-1), \\
                    \sqrt{\frac{\beta(-3\zeta+k+3)}
                             {3(2k+3)\big((1+\beta z_b)^\frac{2(-3\zeta+k+3)}{3(2k+3)}-1\big)}}
                       \left( 1 + \beta |z|\right)^{-\frac{6\zeta  + 4k+ 3}{6(2k+3)}}
                    &  \text{~for solution II}~(\zeta\neq\frac{k+3}{3}).
                   \end{array}
           \right.
             \label{STbrane_scalar_zero_mode}
\end{eqnarray}
It can be seen that, when the extra dimension is finite, the KR zero mode can be localized on the positive tension brane providing that the coupling constant between  the KR and dilaton fields satisfies the condition (\ref{STbrane_KR_norcondition1_11}).
Note that, different from the case of the vector field, the KR zero mode cannot be localized on the positive tension brane anymore if there is no coupling between
the KR and dilaton fields ($\zeta = 0$) for both two solutions.
When the size of extra dimension is infinite,
the KR zero mode can be localized on the positive tension brane
providingl $\zeta>-k-1$ for solution I and $\zeta>\frac{k+3}{3}$ for solution II.

When $\zeta= -k-1$ for solution I and $\zeta= \frac{k+3}{3}$ for solution II, the zero mode $\Omega_{0}(z)$ is
\begin{eqnarray}
 \Omega_{0}(z) =\sqrt{\frac{\beta}{2 \ln(1+\beta z_b)}} \frac{1}{\sqrt{(1+\beta |z|)}}.
 \label{STbrane_special_zeromode_KR}
\end{eqnarray}
It is clear that the KR zero mode for this case is also localized on the positive tension brane.

The $Z_2$ symmetry implies that the KR KK modes should satisfy $\partial_z(e^{-R_i A(z)} \Omega_n)=0$
at $z=0$, with which we get the general solution of Eq. (\ref{STbrane_KR_equation}):
\begin{eqnarray}
 \Omega_n(z)={N}(1+\beta |z|)^\frac{1}{2}
              \Big[ \text{J}_{P_{\text{kr}}^i}(\bar{z}) +
                 \alpha_{\text{kr}}^i ~ \text{Y}_{P_{\text{kr}}^i}(\bar{z})
              \Big],
\end{eqnarray}
where
\begin{eqnarray}
    P_{\text{kr}}^i &\equiv&  {\frac{\sqrt{1-Q_{\text{kr}}^i}}{2}},~~~\\
    Q_{\text{kr}}^I &\equiv& \frac{(-2\zeta+4k+7)(2\zeta-1)}{(2k+3)^2},~~~
    Q_{\text{kr}}^{II} \equiv -\frac{(6\zeta+4k+3)(6\zeta+16k+21)}{9(2k+3)^2},\\
  \alpha_{\text{kr}}^I &\equiv&
       -\frac{\beta \big[1+3k+(P_{\text{kr}}^I-1)(2k+3)-\zeta \big]
          \text{J}_{P_{\text{kr}}^I}( \frac{m_n}{\beta}) -
                (2k+3)m_n \text{J}_{P_{\text{kr}}^I+1}( \frac{m_n}{\beta})}
             {\beta \big[1+3k+(P_{\text{kr}}^I-1)(2k+3)-\zeta \big]
                \text{Y}_{P_{\text{kr}}^I}( \frac{m_n}{\beta})  -
                (2k+3)m_n \text{Y}_{P_{\text{kr}}^I+1}( \frac{m_n}{\beta})},\\
  \alpha_{\text{kr}}^{II} &\equiv&
       {-\frac{\beta \big[3-5k+(P_{\text{kr}}^{II}-1)(6k+9)-3\zeta \big]
                   \text{J}_{P_{\text{kr}}^{II}}(\frac{m_n}{\beta})
                -3(2k+3)m_n \text{J}_{P_{\text{kr}}^{II}+1}(\frac{m_n}{\beta})}
             {\beta\big[3-5k+(P_{\text{kr}}^{II}-1)(6k+9)-3\zeta\big]
                     \text{Y}_{P_{\text{kr}}^{II}}(\frac{m_n}{\beta})
                  -3(2k+3)m_n \text{Y}_{P_{\text{kr}}^{II}+1}(\frac{m_n}{\beta})}}.
\end{eqnarray}

\begin{figure}[htb]
\subfigure[$k=-2$]{
\includegraphics[width=7cm,height=4.5cm]{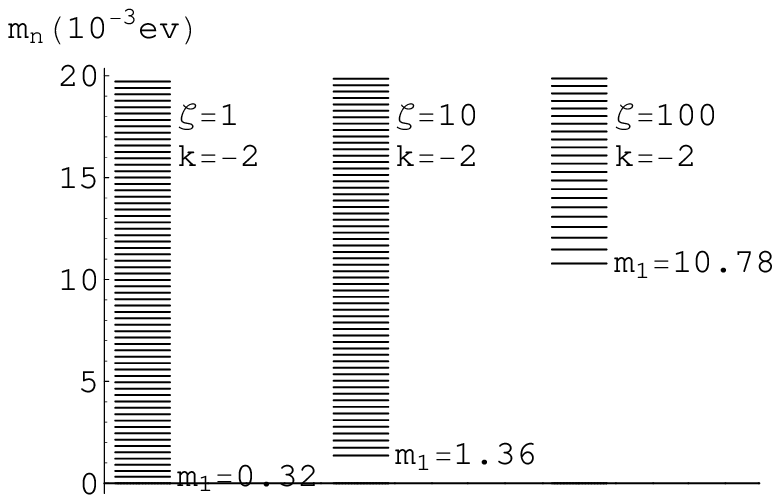}}
\subfigure[$\zeta=10$]{
\includegraphics[width=7cm,height=4.5cm]{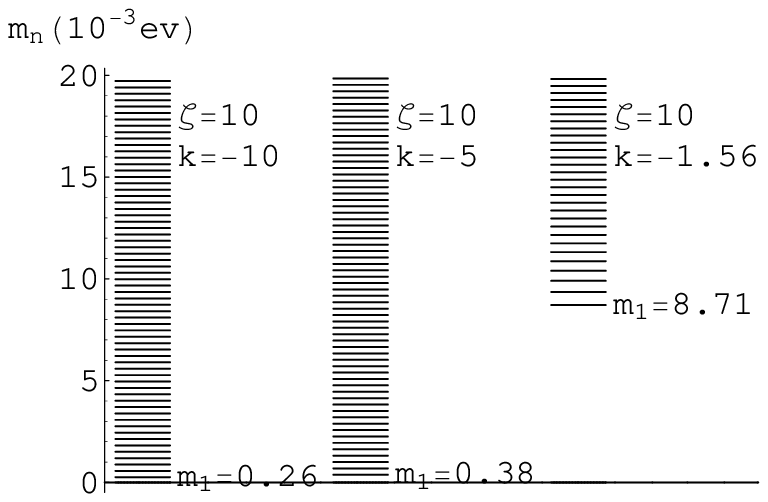}}
 \caption{The mass spectrum of the KR KK modes for solution I.
 The parameters  are set to $\beta=10^{12}$\text{ev} and $z_b=10^4$\text{ev}$^{-1}$.}
 \label{STbrane_fig_Spectrum_KR_SolutionI}
\end{figure}

\begin{figure}[htb]
\subfigure[$k=-1$]{
\includegraphics[width=7cm,height=4.5cm]{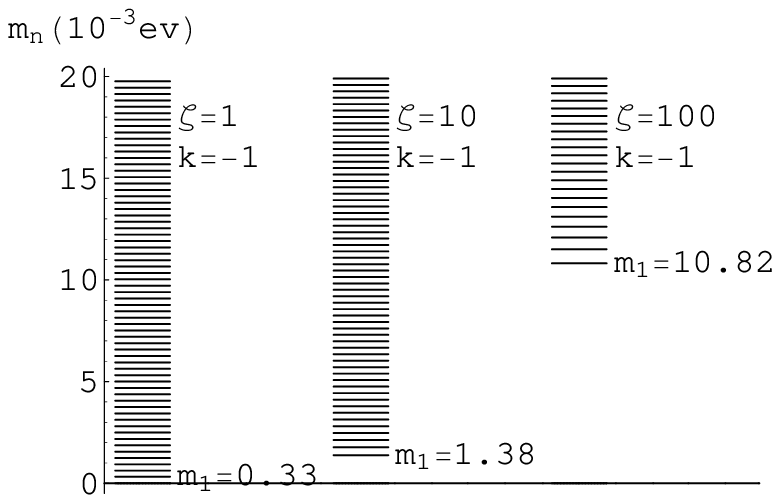}}
\subfigure[$\zeta=10$]{
\includegraphics[width=7cm,height=4.5cm]{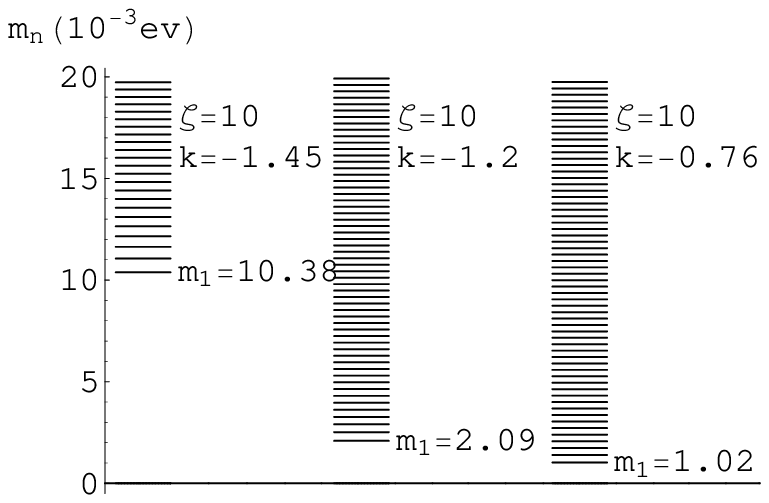}}
 \caption{The mass spectrum of the KR KK modes for solution II.
 The parameters  are set to $\beta=10^{12}$\text{ev} and $z_b=10^4$\text{ev}$^{-1}$.}
 \label{STbrane_fig_Spectrum_KR_SolutionII}
\end{figure}

With another boundary condition $\partial_z(e^{-R_i A(z)} \Omega_n)=0$ at $z=z_b$,
we can obtain the mass spectrum from the following equations
when we consider the light modes in the long range case:
\begin{eqnarray}
  \frac{\text{J}_{P_{\text{kr}}^I} (\bar{z_b}) +\alpha_{\text{kr}}^I \text{Y}_{P_{\text{kr}}^I} (\bar{z_b}) }
      {\text{J}_{P_{\text{kr}}^I+1} (\bar{z_b})+
                        \alpha_{\text{kr}}^I \text{Y}_{P_{\text{kr}}^I+1} (\bar{z_b})  }
    &=&\frac{(2k+3) \bar{z_b}}{1+3k+(P_{\text{kr}}^I-1)(2k+3)-\zeta } ~~~\text {~for solution I},\\
  \frac{\text{J}_{P_{\text{kr}}^{II}} (\bar{z_b}) +\alpha_{\text{kr}}^{II} \text{Y}_{P_{\text{kr}}^{II}} (\bar{z_b}) }
      {\text{J}_{P_{\text{kr}}^{II}+1} (\bar{z_b})+
                        \alpha_{\text{kr}}^{II} \text{Y}_{P_{\text{kr}}^{II}+1} (\bar{z_b})  }
    &=&\frac{3(2k+3) \bar{z_b}}{3-5k+(P_{\text{kr}}^{II}-1)(6k+9)+3\zeta }~\text {~for solution II}.
\end{eqnarray}
The mass spectrum for different values of the parameters is plotted in figs. \ref{STbrane_fig_Spectrum_KR_SolutionI} and \ref{STbrane_fig_Spectrum_KR_SolutionII},
from which we can see that $m_1$ increases with the coupling constant $\zeta$ and the parameter $k$ for solution I, whereas it increases with $\zeta$ but decreases with $k$ for solution II. The mass
gap $\Delta m_n$ approaches a constant for large level $n$.

Particularly, when $\zeta=-k/3$, the effective potentials
\begin{eqnarray}
 V_{\text{kr}}^{I}(z)=V_{\text{kr}}^{II}(z)
             =\frac{7 \beta^2}{36(1+ \beta |z|)^2}-
         \frac{\beta}{3} \Big[\delta(z)-\frac{\delta(z-z_b)}{1+\beta z_b}\Big]
\end{eqnarray}
are the same. This is similar to the cases of scalar and vector fields.
It can be shown that the KR zero mode is localized on the positive tension brane for both solutions.


\subsection{Spin-1/2 fermion field}
\label{STbrane_secFermionFields}

Finally, we turn to investigate localization and mass spectrum
of a spin-1/2 fermion field on the scalar-tensor brane.
In order to localize a fermion on the thick brane generated by an odd scalar field $\phi(z)$,
the Yukawa coupling should be introduced
\cite{Liu0708,Volkas2007,LiuPRD2008,20082009,Fu2011,Liu0803,MelfoPRD2006,0901.3543,Liu0907.0910,KoleyCQG2005,mass5Dscalar,LBCastro1,LBCastro2}.
But when the scalar field $\phi(z)$ is even, the Yukawa coupling (i.e. $\eta\bar\psi F(\phi)\psi$) cannot preserve the
$Z_2$ reflection symmetry of the fermion Lagrangian  and hence cannot ensure localization of the fermion on the brane.
Recently, the authors of Ref. \cite{XuZengGuang} analyzed this problem and
introduced a new localization mechanism to localize the fermion. Following this mechanism, we would like to
analyze localization and spectrum of a fermion on the scalar-tensor brane
by using the new scalar-fermion coupling form $\eta \bar \Psi \Gamma^M \partial_M F(\phi)  \gamma^5 \Psi$.

The Dirac action describing a five-dimensional massless Dirac spinor coupled to the background scalar field $\phi$ reads \cite{XuZengGuang}
\begin{equation}
S_\frac{1}{2} = \int d^5 x  \sqrt{-g}\left [ \bar \Psi \Gamma^M (\partial_M + \omega_M) \Psi
    + \eta \bar \Psi \Gamma^M \partial_M F(\phi)  \gamma^5 \Psi \right],
\end{equation}
where $\eta$ is the coupling constant. With the metric (\ref{STbrane_linee}), the explicit five-dimensional Dirac equation reads
\begin{eqnarray}
 \left[\gamma^{\mu}\partial_{\mu}
         + \gamma^5 \left(\partial_z  +2 \partial_{z} A \left( z \right) \right)
         -\eta \partial_z F(\phi)
 \right ] \Psi =0. \label{STbrane_DiracEq11}
\end{eqnarray}
In order to solve the above equation, we
make the following general chiral decomposition for the Dirac spinor
\begin{eqnarray}
 \Psi(x,z) &=& \sum_n\psi_{Ln}(x) \hat{f}_{Ln}(z)
 +\sum_n\psi_{Rn}(x) \hat{f}_{Rn}(z) \nonumber \\
 &=&\sum_n\psi_{Ln}(x) f_{Ln}(z) a^{-2}(z)
 +\sum_n\psi_{Rn}(x) f_{Rn}(z) a^{-2}(z), \label{KKdecomposition}
\end{eqnarray}
where $\hat{f}_{Ln,Rn}(z)=a^{-2}(z){f}_{Ln,Rn}(z)$, and
the left-handed and
right-handed components of a four-dimensional Dirac field satisfy $\gamma^5 \psi_{Ln,Rn}(x)=\mp\psi_{Ln,Rn}(x)$. 
Then substituting the decomposition (\ref{KKdecomposition}) into the Dirac equation (\ref{STbrane_DiracEq11}), we find that $\psi_{Ln,Rn}(x)$ satisfy the four-dimensional massive Dirac equations
$\gamma^{\mu}\partial_{\mu}\psi_{Ln,Rn}(x)=m_n\psi_{Rn,Ln}(x)$, and the left- and right-handed KK modes $f_{Ln,Rn}(z)$ satisfy the following coupled
equations
\begin{eqnarray}
 \left[\partial_z \mp \eta F'(\phi) \right]f_{Ln,Rn}(z)
  = \pm m_n f_{Rn,Ln}(z),\label{STbrane_CoupleEq}
\end{eqnarray}
from which, we get the Schr\"{o}dinger-like
equations for the fermion KK modes
\begin{eqnarray}
\label{STbrane_SchEqFermion}
  \big(-\partial^2_z + V_L^i(z) \big)f_{Ln}
            &=&m_n^2 f_{Ln},
   \label{STbrane_SchEqLeftFermion}  \\
  \big(-\partial^2_z + V_R^i(z) \big)f_{Rn}
            &=&m_n^2 f_{Rn},    ~~~(i=I,II)
   \label{STbrane_SchEqRightFermion}
\end{eqnarray}
with the effective potentials given by \cite{XuZengGuang}
\begin{subequations}
\begin{eqnarray}
  V_L^i(z)&=& \eta^2\big(\partial_z F(\phi)\big)^2
    + \eta\partial_z^2 F(\phi), \label{STbrane_VL}\\
  V_R^i(z)&=&   V_L(z)|_{\eta \rightarrow -\eta}. \label{STbrane_VR}
\end{eqnarray}\label{STbrane_Vfermion}
\end{subequations}
In order to obtain the effective four-dimensional Dirac action for the massless chiral fermion and massive fermions:
\begin{eqnarray}
 S_{1/2} &=& \int d^5 x \sqrt{-g} ~\bar{\Psi}
     \left[  \Gamma^M (\partial_M+\omega_M)
     + \eta \Gamma^M \partial_M F(\phi)  \gamma^5 \right] \Psi  \nn \\
  &=& \sum_{n}\int d^4 x \left(~\bar{\psi}_{Rn}
      \gamma^{\mu}\partial_{\mu}\psi_{Rn}
        -~\bar{\psi}_{Rn}m_{n}\psi_{Ln} \right) 
  +\sum_{n}\int d^4 x \left(~\bar{\psi}_{Ln}
      \gamma^{\mu}\partial_{\mu}\psi_{Ln}
        -~\bar{\psi}_{Ln}m_{n}\psi_{Rn} \right)  \nn \\
  &=&\sum_{n}\int d^4 x
    ~\bar{\psi}_{n}
      (\gamma^{\mu}\partial_{\mu} -m_{n})\psi_{n},
\end{eqnarray}
we should introduce the following orthonormality conditions:
\begin{eqnarray}
 \int_{-\infty}^{\infty} f_{Lm} f_{Ln}dz=\delta_{mn}
   = \int_{-\infty}^{\infty} f_{Rm} f_{Rn}dz,~~
 \int_{-\infty}^{\infty} f_{Lm} f_{Rn}dz=0.
 \label{STbrane_orthonormality}
\end{eqnarray}
It can be seen from Eq. (\ref{STbrane_Vfermion}) that, in order to trap the four-dimensional fermions on the positive tension brane, some kind of scalar-fermion coupling should be introduced ($\eta \neq 0$), and
the effective potential $V_L^{i}(z)$ or $V_R^{i}(z)$ should have a minimum
at the location of the brane.

In Ref. \cite{XuZengGuang}, the authors took $F(\phi(z))=\phi^{q}(z)$ with $q$ an integral and found the following result: for $q=1$, if $\eta>-(3+2k)/4$ for solution I and $\eta>(3+2k)/4$ for solution II, the zero mode of the left-handed fermion can be localized on the positive tension brane; for odd $q> 1$ and even $q>1$, the zero modes of
the left- and right-handed fermions are respectively localized on the positive tension branes when $\eta>0$.
Here, considering that the scalar $\phi$ is a dilaton, we take its  exponential function as a new kind of coupling, i.e.,
\begin{eqnarray}
F(\phi(z))=e^{v \phi},  \label{F_phi}
\end{eqnarray}
and investigate localization of the fermion field. We will find that the similar result is that only one of the left and right-handed fermion zero mode can be localized on the positive tension brane, while the difference is that we will obtain discrete mass spectrum for some range of the parameters with the new coupling (\ref{F_phi}).

\begin{figure}[htb]
\begin{center}
\subfigure[$\eta=0.5, v=0.2$]{\label{STbrane_fig_fermion_potentialV_gr1}
\includegraphics[width=4.5cm,height=3.7cm]{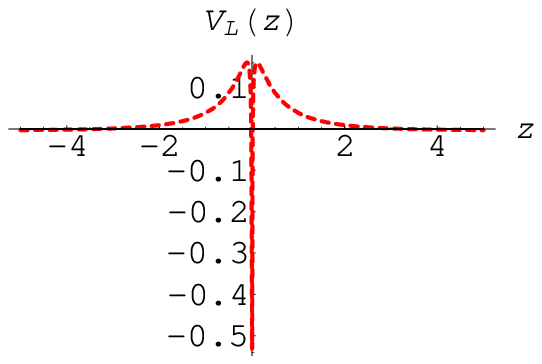}}
\subfigure[$\eta=0.5, v=-2$]{\label{STbrane_fig_fermion_potentialV_gr3}
\includegraphics[width=4.5cm,height=3.7cm]{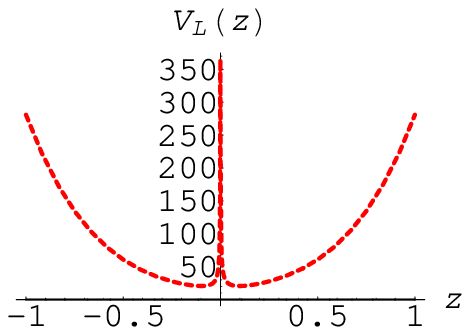}}
\subfigure[$\eta=-0.5, v=-0.8$]{\label{STbrane_fig_fermion_potentialV_gr5}
\includegraphics[width=4.5cm,height=3.7cm]{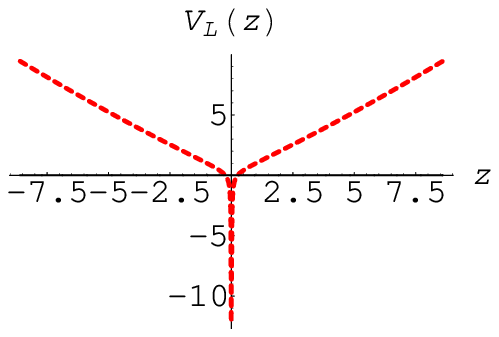}}
\end{center}
\begin{center}
\subfigure[$\eta=0.5, v=0.2$]{\label{STbrane_fig_fermion_potentialV_gr2}
\includegraphics[width=4.5cm,height=3.7cm]{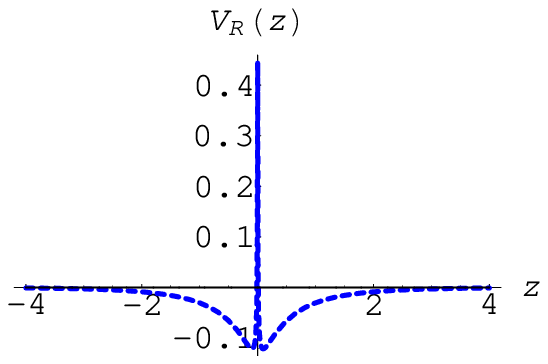}}
\subfigure[$\eta=0.5, v=-2$]{\label{STbrane_fig_fermion_potentialV_gr4}
\includegraphics[width=4.5cm,height=3.7cm]{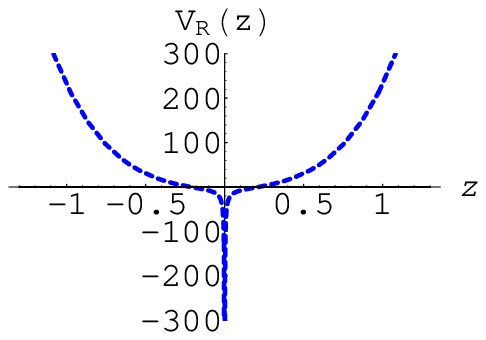}}
\subfigure[$\eta=-0.5, v=-0.8$]{\label{STbrane_fig_fermion_potentialV_gr6}
\includegraphics[width=4.5cm,height=3.7cm]{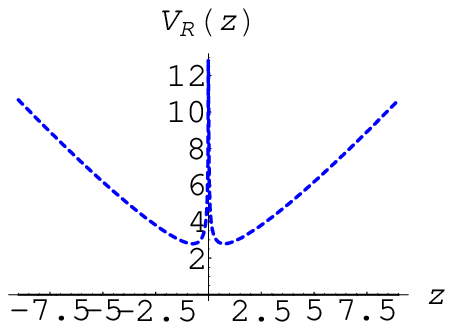}}
\end{center}
 \caption{The shapes of the effective potentials of the left- and the right-handed fermion $kk$ modes for solution I.
 The parameter $k$ is set to $k=-2$.
 }
\label{STbrane_fig_fermion_potentialV_grs}
\end{figure}

The explicit forms of the effective potentials (\ref{STbrane_Vfermion}) read as
\begin{eqnarray} \label{STbrane_VLVRfermion_explicit1}
 V_L^{i}(z) &=& \frac{2v \eta\beta^2}{(2k+3)^2}(1+ \beta |z|)^{\pm\frac{2v}{2k+3}-2}
        \left[ 2v\pm(-2k-3)+ 2v\eta(1+\beta |z|)^{\pm\frac{2v}{2k+3}} \right]  \nonumber\\
      && \pm{\frac{4v \eta \beta }{2k+3}}\Big[ \delta(z)- \frac{1}{1+\beta z_b} \delta(z-z_b) \Big], \\
  V_R^i(z) &=& V_L^i(z)|_{\eta \rightarrow -\eta} ~~~~~~(i=I,II),
\end{eqnarray}
where $+$ for solution I and $-$ for solution II in this section.
When  $v = (2k+3)/2$ for solution I and $v = -(2k+3)/2$ for solution II, the left- and right-handed potentials in the bulk are the same constants:
\begin{eqnarray}
 V_{L,R}^i(z)=\eta^2 \beta^2 .
\end{eqnarray}
The values of $V_{L}^i(z)$ at $z = 0$  are
\begin{eqnarray}
 V_L^i(0)&=& \frac{2v \eta \beta^2  } {(2k+3)^2}
               \big[2v \pm(-2k-3)+ 2v \eta \big]
            \pm \frac{4v \eta \beta}{2k+3} \delta(0).
\end{eqnarray}
The shapes of the potentials are shown
in fig. \ref{STbrane_fig_fermion_potentialV_grs}.

%
%
%

The solutions of the left- and right-handed fermion zero modes are
\begin{eqnarray}
 f_{L0}^i(z)& \propto&
          \exp\left(+\eta F(\phi)\right)=e^{+\eta(1+\beta |z|)^{\pm\frac{2v}{2k+3}}}
        \label{STbrane_zeroModefL0},~~\\
 f_{R0}^i(z) &\propto&
        \exp\left(-\eta F(\phi)\right)=e^{-\eta(1+\beta |z|)^{\pm\frac{2v}{2k+3}}}.
          \label{STbrane_zeroModefR0}
\end{eqnarray}
It is clear that the left- and right-handed fermion zero modes
cannot be localized on the positive {tension} brane at the same time.
The left-handed fermion zero mode can be localized on the positive {tension} brane
{if $\eta v>0$ and the extra dimension is finite}.
In order to check whether the left-handed zero mode $f_{L0}^i(z)$ can be localized
when the extra dimension is infinite,
we need to consider the following normalization conditions:
\begin{eqnarray}
 \int_{-\infty}^{\infty} dz {~f_{L0}^i}^2(z) \propto
 \int_{-\infty}^{\infty}e^{2\eta(1+\beta |z|)^{\pm\frac{2v}{2k+3}}} dz <\infty .
\end{eqnarray}\label{STbrane_eq:Norconditon-z}
The condition is turned out to be
\begin{eqnarray}
 \eta<0, ~~v<0.\label{STbrane_conditionCaseI}
\end{eqnarray}
If the extra dimension is finite, then the right-handed fermion zero mode is localized
on the negative tension brane under the condition {$\eta v>0$}.

\begin{figure}[htb]
\begin{center}
\subfigure[$\tan(m_n z_b)$ and $\frac{(2k+3)}{2}(m_n z_b)$ for solution I.]{\label{STbrane_Fig_fermion_similarSolution_I}
\includegraphics[width=10cm]{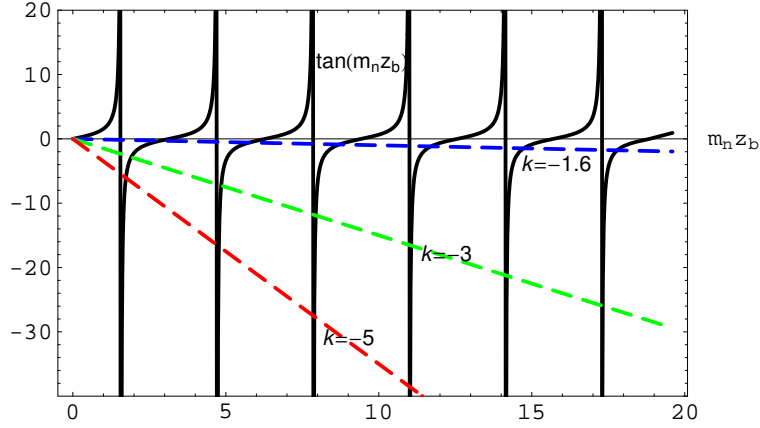}}
\subfigure[$\tan(m_n z_b)$ and $\frac{3(2k+3)}{2(4k+3)}(m_n z_b)$ for solution II.]{\label{STbrane_Fig_fermion_similarSolution_II}
\includegraphics[width=10cm]{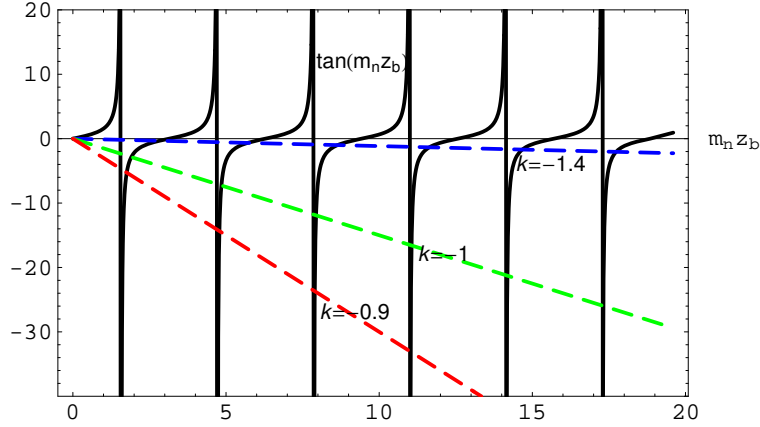}}
\end{center}
 \caption{The functions $\tan(m_n z_b)$ (black lines), $\frac{(2k+3)}{2}(m_n z_b)$ (dashed lines in the left subfigure) and $\frac{3(2k+3)}{2(4k+3)}(m_n z_b)$ (dashed lines in the right subfigure).
 The parameters  are set to $\beta=10^{12}$\text{eV} and $z_b=10^4$\text{eV}$^{-1}$.}
\label{STbrane_Fig_fermion_similarSolution}
\end{figure}

Next, we consider the massive fermion KK modes for the case of finite extra dimension.
For simplicity, we only consider a free fermion, which means that the coupling constant
is set to zero ($\eta=0$). The boundary conditions are decided by the $Z_2$ symmetry:
$\partial_z (a^{-2}(z) f_{L,R}(z))|_{z=0,z_b}=0$.
The general solutions of the massive fermion KK modes are
\begin{eqnarray}
 f_{Ln,Rn}^I(z) &=& {N} \Big(\cos(m_n z)+\frac{2\beta}{(2k+3)m_n}\sin(m_n z) \Big) ,\\
 f_{Ln,Rn}^{II}(z) &=&{N} \Big(\cos(m_n z)+\frac{2(4k+3)\beta}{3(2k+3)m_n}\sin(m_n z) \Big) .
\end{eqnarray}
With the boundary condition at $z=z_b$,
the exact spectrum is determined by the following equation:
\begin{eqnarray}
 \tan(m_n z_b)&=&\frac{2(2k+3)m_n z_b }{4 +(2k+3)^2 (1+\beta z_b) \frac{ m_n^2}{\beta^2}} ~~~~~~~~~~~~~~~~~\text{for solution I},\label{STbrane_Eq_massivefermion_modes_explicitI}\\
 \tan(m_n z_b)&=&\frac{6 {(4k+3)(2k+3)}m_n z_b }
              {4(4k+3)^2 +9(2k+3)^2(1+\beta z_b ) \frac{m_n^2}{\beta^2} }  ~~~~\text{for solution II}.
 \label{STbrane_Eq_massivefermion_modes_explicitII}
\end{eqnarray}
{For those fermion KK modes satisfying
$m_n  \ll 1$eV, we have $\frac{m_n^2}{\beta^2} \ll  10^{-24}$ and
$(1+\beta z_b) \frac{m_n^2}{\beta^2} \ll 10^{-8}$.
If $|k|< 10^4$ for solution I and
$\left|\frac{2k+3}{4k+3}\right| < \frac{2}{3}\times 10^4$ (i.e., $k<-\frac{3.00045}{4.00030}$) for solution II, the approximative mass spectrum can be determined by
\begin{eqnarray}
  \tan(m_n z_b)&=&\frac{(2k+3)}{2} m_n z_b  ~~~~~~~~\text{for solution I}, \nonumber\\
  \tan(m_n z_b)&=&\frac{3(2k+3)}{2(4k+3)} m_n z_b  ~~~~~~~\text{for solution II}.
\end{eqnarray}
We have plotted the functions
$\tan(m_n z_b)$, $\frac{(2k+3)}{2}(m_n z_b)$, and $\frac{3(2k+3)}{2(4k+3)}(m_n z_b)$ in fig.~\ref{STbrane_Fig_fermion_similarSolution},
from which one can see that when
\begin{eqnarray}
  \begin{array}{ll}
    -10^4  \lesssim k< -3                   & ~~{\text{for solution I and}}  \\
    -1 ~~~\lesssim k <-\frac{3.00045}{4.00030} & ~~{\text{for solution II}},
  \end{array}\label{AppCondtion}
\end{eqnarray}
the fermion mass spectrum for both solutions can be given by
\begin{eqnarray}
 m_n= \frac{(2n-1)\pi}{2z_b}=\frac{(2n-1)\pi}{2}\times 10^{-4} \text{eV}. ~~(n=1,2,\cdots) \label{STbrane_Eq_massivefermion_modes_approximative}
\end{eqnarray}
}

\begin{figure}[htb]
\subfigure[Exact spectrum]{\label{STbrane_fig_Spectrum_massive_fermion_SolutionIa}
\includegraphics[width=7cm,height=4.5cm]{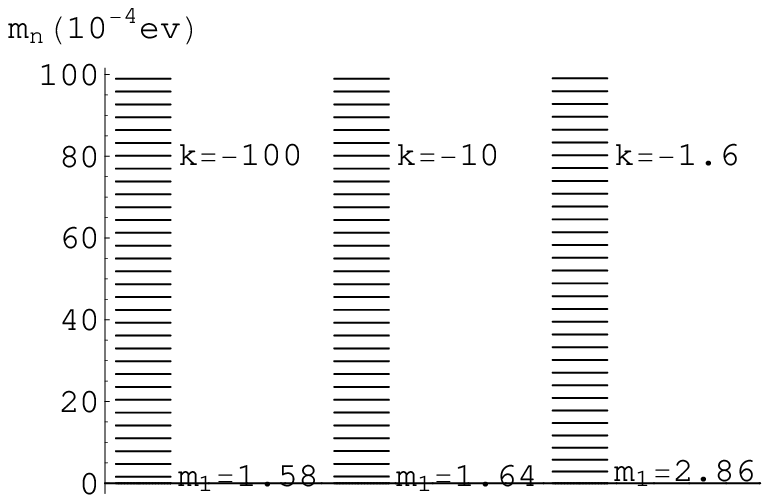}}
\subfigure[Exact and  approximative spectra]{ \label{STbrane_fig_Spectrum_massive_fermion_SolutionIb}
\includegraphics[width=7cm,height=4.5cm]{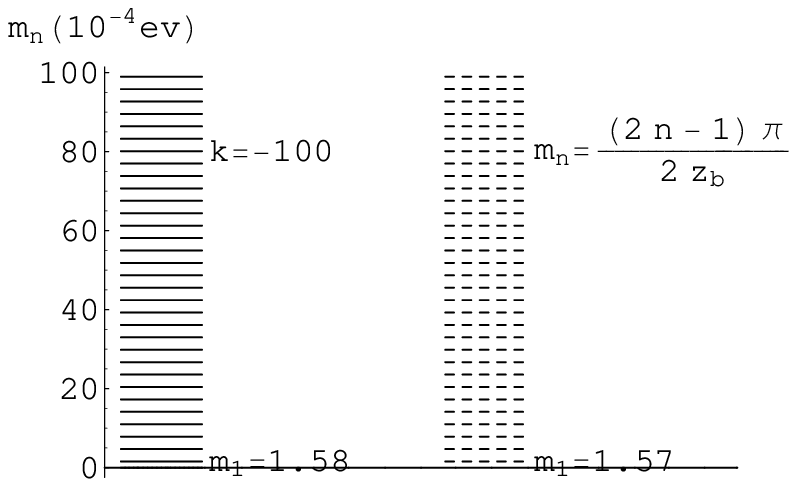}}
 \caption{The exact (solid) and  approximative (dashing) spectra of the free fermion
 for solution I. The parameters $\beta$ and $z_b$ are set to $\beta=10^{12}$\text{ev} and $z_b=10^4$\text{ev}$^{-1}$.}
 \label{STbrane_fig_Spectrum_massive_fermion_SolutionI}
\end{figure}

\begin{figure}[htb]
\subfigure[Exact spectrum]{\label{STbrane_fig_Spectrum_massive_fermion_SolutionIIa}
\includegraphics[width=7cm,height=4.5cm]{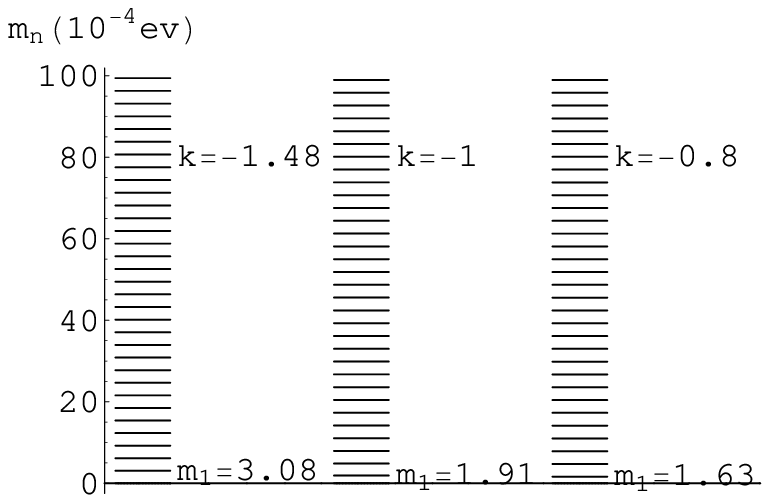}}
\subfigure[Exact and  approximative spectra]{ \label{STbrane_fig_Spectrum_massive_fermion_SolutionIIb}
\includegraphics[width=7cm,height=4.5cm]{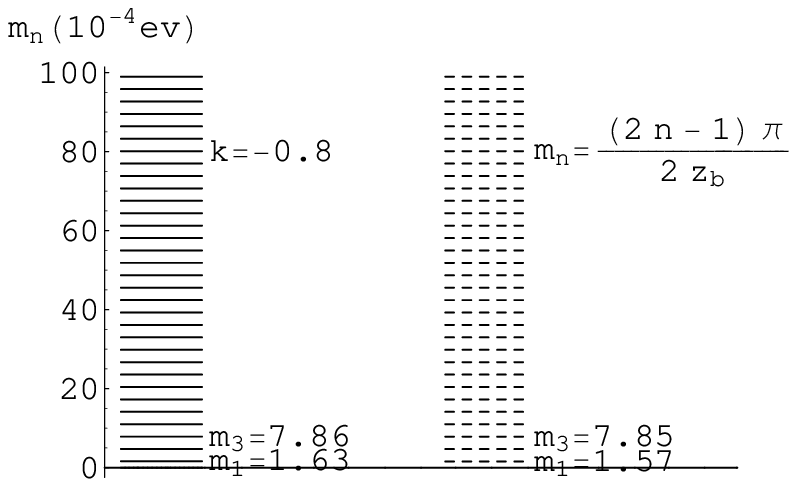}}
 \caption{The exact (solid) and  approximative (dashing) spectra of the free fermion
 for solution II. The parameters $\beta$ and $z_b$ are set to $\beta=10^{12}$\text{ev} and $z_b=10^4$\text{ev}$^{-1}$.}
 \label{STbrane_fig_Spectrum_massive_fermion_SolutionII}
\end{figure}

The mass spectrum numerically calculated from Eqs. (\ref{STbrane_Eq_massivefermion_modes_explicitI}),
(\ref{STbrane_Eq_massivefermion_modes_explicitII})
and the approximative one given in (\ref{STbrane_Eq_massivefermion_modes_approximative})
are plotted in figs.~\ref{STbrane_fig_Spectrum_massive_fermion_SolutionI} and ~\ref{STbrane_fig_Spectrum_massive_fermion_SolutionII}.
From figs.~\ref{STbrane_fig_Spectrum_massive_fermion_SolutionIa}
and ~\ref{STbrane_fig_Spectrum_massive_fermion_SolutionIIa},
we reach the conclusion that the mass of the first massive fermion KK mode $m_1$
increases and decreases with the parameter $k$ for solutions I and II, respectively.
From figs.~\ref{STbrane_fig_Spectrum_massive_fermion_SolutionIb} and ~\ref{STbrane_fig_Spectrum_massive_fermion_SolutionIIb},
we see that the approximative analytical spectrum is consistent with the exact numerical one under the condition (\ref{AppCondtion}).


\section{Discussions and conclusions}\label{STbrane_secConclusion}

The scalar-tensor braneworld model presented in Ref. \cite{YangKe} not only solves
the gauge hierarchy problem but also reproduces a correct Friedmann-like
equation on the brane, and so overcomes the cosmological problem in the Randall-Sundrum model.
In this model, there are two similar but different brane solutions.
In each solution, there are two branes, one with positive tension and another with negative tension.
Our world is confined on the positive tension brane.
The tensor perturbation of the brane system is stable
and the mass spectra of the gravitational KK modes for both brane solutions are the same.
Therefore, one cannot distinguish the two solutions by the mass spectra of gravity.

In this paper, we investigated localization of the zero modes and mass spectra for
various bulk matter fields (i.e., scalar, vector, KR, and fermion fields) on the scalar-tensor  brane.
For the scalar, vector, and KR fields, we considered their interaction with the background scalar field
(the dilaton $\phi$) that generates the brane.
For the fermion, following Ref.~\cite{XuZengGuang},
we introduced a new scalar-fermion coupling $\bar \Psi \Gamma^M \partial_M e^{v\phi} \gamma^5 \Psi$
instead of the usual Yukawa coupling for the reason of the even parity of the dilaton $\phi$.
We found that the mass spectra of each bulk matter field for the two brane solutions are different,
which implies that the two brane solutions are not physically equivalent.

It was found that the zero modes of various bulk matter fields
can be localized on the positive tension brane for the two solutions under some conditions,
which are collected in Table \ref{table_localization_condition}.
It can be seen that the localization conditions for the case of infinite extra dimension
are stronger than the case of finite extra dimension. When extra dimension is finite,
the scalar and vector zero modes can be localized on the positive tension brane
even if there is no interaction with the background scalar field,
while the KR and fermion zero modes cannot be localized if there is no interaction.

\begin{table}
\begin{center}
\caption{Localization conditions for the zero modes of various bulk matter fields. Here $z_b$ is
the size of the extra dimension.
}\label{table_localization_condition}
\begin{tabular}{|c|c|c|c|c|}
  \hline
  Bulk matter & Lagrangian & $~~~~z_b~~~$ & ~Solution I~  & ~Solution II~ \\\hline \hline
  \multicolumn{1}{|c|}{\multirow {2}{*}{Scalar}}
  &\multicolumn{1}{|c|}{\multirow {2}{*}{$\mathcal{L}_0= -\frac{1}{2} e^{\lambda\phi} \partial_M \Phi \partial^M \Phi$}}
           &  finite
           & $\lambda >-\frac{3}{2}$
           & $\lambda >\frac{4k+3}{2}$ \\
           \cline{3-5}
      &  &  infinite
           & $\lambda >-k-3$
           & $\lambda >3k+3$ \\\hline
  \multicolumn{1}{|c|}{\multirow {2}{*}{Vector}}
  &\multicolumn{1}{|c|}{\multirow {2}{*}{$\mathcal{L}_1= -\frac{1}{4}  e^{\tau\phi} F_{MN}F^{MN}$}}
           &  finite
           & $\tau >-\frac{1}{2}$
           & $\tau >\frac{4k+3}{6}$ \\
           \cline{3-5}
      &  &  infinite
           & $\tau >-k-2$
           & $\tau >\frac{5k+6}{3}$ \\\hline
  \multicolumn{1}{|c|}{\multirow {2}{*}{KR}}
  &\multicolumn{1}{|c|}{\multirow {2}{*}{$\mathcal{L}_{\text{kr}}=- e^{\zeta\phi} H_{MNL} H^{MNL}$}}
           &  finite
           & $\zeta>\frac{1}{2}$
           & $\zeta>-\frac{4k+3}{6}$ \\
           \cline{3-5}
      &  &  infinite
           & $\zeta>-k-1$
           & $\zeta>-\frac{k+3}{3}$ \\\hline
  \multicolumn{1}{|c|}{\multirow {2}{*}{Fermion}}
  &\multicolumn{1}{|c|}{\multirow {2}{*}{$\mathcal{L}_{1/2}=\bar \Psi \Gamma^M \left [  D_{M}
              + \eta \partial_M e^{v\phi}  \gamma^5 \right] \Psi$}}
           &  finite
           & $\eta v >0$
           & $\eta  v >0$ \\
           \cline{3-5}
      &  &  infinite
           & $\eta<0, v<0$
           & $\eta<0, v<0$ \\
  \hline
\end{tabular}
\end{center}
\end{table}

The bound massive KK modes and discrete mass spectra were also obtained for the case of finite extra dimension. For the scalar, vector, and KR fields, we get the analytical solutions of their KK modes by solving the corresponding Schr\"{o}dinger-like equations, and the numerical mass spectra by considering the boundary conditions. It was found that
the mass of the first massive KK mode $m_1$
increases with the coupling constant and the parameter $k$ for
solution I, while it increases with the coupling constant but
decreases with the parameter $k$ for
solution II. The mass gap between two adjoining KK modes becomes smaller and smaller with the increase of the level $n$ and then trends to a constant.

For the fermion field, we did not find the analytic solution of the fermion KK modes when there exists coupling because the effective potential is too complex.
Therefore, we only considered the free fermion and got the approximate analytical mass spectrum.
It was found that the numerical
$m_1$ slowly increases and decreases with the parameter $k$ for
solution I and solution II, respectively.
The mass spectrum approaches equidistant for the higher excited states.

\section{Acknowledgement}
This work was supported by the National Natural Science Foundation of China (Grants No. 11075065 and No. 11375075) and
the Fundamental Research Funds for the Central Universities (No. lzujbky-2013-18).


\begin{thebibliography}{99}

 \bibitem{RSmodel1}
 L. Randall and R. Sundrum,
    Phys. Rev. Lett. \textbf{83} (1999) 3370,
    arXiv:hep-ph/9905221.

\bibitem{RSmodel2}
 L. Randall and R. Sundrum,
    Phys. Rev. Lett. \textbf{83} (1999) 4690,
    arXiv:hep-th/9906064.

\bibitem{Hatanaka1999}
 H. Hatanaka, M. Sakamoto, M. Tachibana, and K. Takenaga,
    Prog. Theor. Phys. \textbf{102} (1999) 1213,
    arXiv:hep-th/9909076.

\bibitem{Das2008}
 S. Das, D. Maity, and S. SenGupta,
    JHEP \textbf{0805} (2008) 042,
    arXiv:0711.1744[hep-th].

\bibitem{Lepe2008}
 S. Lepe, F. Pe\~{n}a, and J. Saavedra,
    Phys. Lett. \textbf{B 662} (2008) 217,
    arXiv:0803.0518[hep-th].

\bibitem{Zhong2010Scalar-Kinetic}
 Y.-X. Liu, Y. Zhong, and K. Yang,
    Europhys. Lett. \textbf{90} (2010) 51001,
    arXiv:0907.1952[hep-th].

\bibitem{Yang2010Weyl}
 Y.-X. Liu, K. Yang, and Y. Zhong,
    JHEP \textbf{1010} (2010) 069,
    arXiv:0911.0269[hep-th].


\bibitem{Balcerzak20011}
 A. Balcerzak and M. P. Dabrowski,
    Phys. Rev. \textbf{D 84} (2011) 063529,
    arXiv:1107.3048[hep-th].

\bibitem{Ahmed20014}
 A. Ahmed, L. Dulny, and B. Grzadkowski,
    Eur. Phys. J. C \textbf{74} (2014) 2862,
    arXiv:1312.3577[hep-th].

\bibitem{Aros2013}
 R. Aros and M. Estrada,
    Phys. Rev. \textbf{D 88} (2013) 027508,
    arXiv:1212.0811[hep-th].

\bibitem{Fu1407.6107}
 Q.-M. Fu, L. Zhao, K. Yang, B.-M. Gu, and Y.-X. Liu,
    Phys. Rev. \textbf{D 90} (2014) 104007,
    arXiv:1407.6107[hep-th].

\bibitem{WangBin2014}
 D.-C. Dai, D. Stojkovic, B. Wang, and C.-Y. Zhang,
    Phys. Rev. \textbf{D90} (2014)064031,
    arXiv:1409.5139[hep-th].

\bibitem{Bazeia1411.0897}
 D. Bazeia, L. Losano, R. Menezes, Gonzalo J. Olmo, D. Rubiera-Garcia,
    {\em Thick brane in $f(R)$ gravity with Palatini dynamics},
    arXiv:1411.0897[hep-th].


\bibitem{Arkani-Hamed1998}
  N.~Arkani-Hamed, S.~Dimopoulos, and G.~R. Dvali,
  Phys. Lett. \textbf{B 429} (1998) 263,
  arXiv:hep-ph/9803315.

\bibitem{Bajc2000}
B. Bajc and G. Gabadadze,
   Phys. Lett. \textbf{B 474} (2000) 282,
   arXiv: hep-th/9912232.

\bibitem{Fuchune}
 C.-E. Fu, Y.-X. Liu, and H. Guo,
 Phys. Rev. \textbf{D 84} (2011) 044036,
 arXiv:1101.0336[hep-th].



\bibitem{Oda2000}
   I. Oda,
   Phys. Lett. \textbf{B 496} (2000) 113,
   arXiv:hep-th/0006203.

\bibitem{Liu0708}
 Y.-X. Liu, X.-H. Zhang, L.-D. Zhang, and Y.-S. Duan,
    JHEP \textbf{0802} (2008) 067,
    arXiv:0708.0065[hep-th].

\bibitem{LiuPRD2008}
  Y.-X. Liu, L.-D. Zhang, L.-J. Zhang, and Y.-S. Duan,
    Phys. Rev. \textbf{D 78} (2008) 065025,
    arXiv:0804.4553[hep-th].

\bibitem{Massimo2002}
  M. Giovannini,
  Phys. Rev. \textbf{D 65} (2002) 124019,
 arXiv: hep-th/0204235.

\bibitem{ZhaoZHH2014}
 Z.-H. Zhao, Q.-Y. Xie, and Y. Zhong,
    {\em New localization mechanism of $U(1)$ gauge vector field on flat branes in (asymptotic) $AdS_{5}$ spacetime},
    arXiv:1406.3098[hep-th].

\bibitem{Carlos2014}
  C. A. Vaquera-Araujo and O. Corradini,
    {\em Localization of Abelian Gauge Fields on Thick Branes},
    arXiv:1406.2892[hep-th].


\bibitem{Cruz2009}
  W. T. Cruz, M. O. Tahim, and C. A. S. Almeida,
  Europhys. Lett. 88 (2009) 41001,
  arXiv:0912.1029[hep-th].


\bibitem{Christiansen2010}
  H. R. Christiansen, M. S. Cunha, and M. O. Tahim,
   Phys. Rev. \textbf{D 82} (2010) 085023,
   arXiv:1006.1366[hep-th].

\bibitem{Christ2012}
 H.~Christiansen and M.~Cunha,
    Eur. Phys. J. \textbf{C 72} (2012) 1942,
    arXiv:1203.2172[hep-th].

\bibitem{CruzWT}
   W. T. Cruz, R. V. Maluf, and C. A. S. Almeida,
  Eur. Phys. J. C \textbf{73} (2013) 2523,
  arXiv:1303.1096[hep-th].


\bibitem{DuYunZhi}
  Y.-Z. Du, L. Zhao, Y. Zhong, C.-E. Fu, and H. Guo,
  Phys. Rev. \textbf{D 88} (2013) 024009,
  arXiv:1301.3204[hep-th].

\bibitem{Volkas2007}
 T. R. Slatyer and R. R. Volkas,
    JHEP \textbf{0704} (2007) 062,
    arXiv:hep-ph/0609003.

\bibitem{RandjbarPLB2000}
  S. Randjbar-Daemi and M. Shaposhnikov,
     Phys. Lett. \textbf{B 492} (2000) 361,
     arXiv:hep-th/0008079.



\bibitem{GuoHeng2013}
 H. Guo, A. Herrera-Aguilar, Y.-X. Liu, D. Malag\'{o}n-Morej\'{o}n, and R. R. Mora-Luna,
    Phys. Rev. \textbf{D 87} (2013) 095011,
    arXiv:1103.2430[hep-th].

\bibitem{Csaki1999}
C.~Csaki, M.~Graesser, C.~F. Kolda, and J.~Terning,
    {Phys. Lett.} \textbf{B 462} (1999) 34,
    arXiv:hep-ph/9906513.

\bibitem{Cline1999}
J.~M. Cline, C.~Grojean, and G.~Servant,
    Phys. Rev. Lett. \textbf{83} (1999) 4245,
    arXiv:hep-ph/9906523.

\bibitem{Shiromizu2000}
T.~Shiromizu, K.~I. Maeda, and M.~Sasaki,
    Phys. Rev. \textbf{D 62} (2000) 024012,
    arXiv:gr-qc/9910076.

\bibitem{YangKe}
 K. Yang, Y.-X. Liu, Y. Zhong, X.-L. Du, and S.-W. Wei,
 Phys. Rev. \textbf{D 86} (2012) 127502,
 arXiv:1212.2735[hep-th].

\bibitem{XuZengGuang}
 Y.-X. Liu, Z.-G. Xu, F.-W. Chen, and S.-W. Wei,
    Phys. Rev. \textbf{D 89} (2014) 086001,
    arXiv:1312.4145[hep-th].

\bibitem{20082009}
 D. Bazeia, F. A. Brito, and R. C. Fonseca,
    Eur. Phys. J. \textbf{C 63} (2009) 163,
    arXiv:0809.3048[hep-th].

\bibitem{Fu2011}
 C.-E. Fu, Y.-X. Liu, and H.~Guo,
    Phys. Rev. \textbf{D 84} (2011) 044036,
    arXiv:1101.0336[hep-th].

\bibitem{Liu0803}
 Y.-X. Liu, L.-D. Zhang, S.-W. Wei, and Y.-S. Duan,
    JHEP \textbf{0808} (2008) 041,
    arXiv:0803.0098[hep-th].



\bibitem{MelfoPRD2006}
 A. Melfo, N. Pantoja, and J. D. Tempo,
    Phys. Rev. \textbf{D 73} (2006) 044033,
    arXiv:hep-th/0601161.

\bibitem{0901.3543}
 C. A. S. Almeida, R. Casana, M. M. Ferreira, and A. R. Gomes,
    Phys. Rev. \textbf{D 79} (2009) 125022,
    arXiv:0901.3543[hep-th].



\bibitem{Liu0907.0910}
 Y.-X. Liu, C.-E. Fu, L. Zhao, and Y.-S. Duan,
    Phys. Rev. \textbf{D 80} (2009) 065020,
    arXiv:0907.0910[hep-th].




\bibitem{KoleyCQG2005}
 R. Koley and S. Kar,
    Class. Quant. Grav. \textbf{22} (2005) 753,
    arXiv:hep-th/0407158.

\bibitem{mass5Dscalar}
R. Davies and D. P. George,
    Phys. Rev. \textbf{D 76} (2007) 104010,
    arXiv:0705.1391[hep-th].

\bibitem{LBCastro1}
L.B. Castro,
    Phys. Rev. \textbf{D 83} (2011) 045002,
    arXiv:1008.3665[hep-th].


\bibitem{LBCastro2}
L.B. Castro and L.E. Arroyo Meza,
    Europhys. Lett. \textbf{102} (2013) 21001,
    arXiv:1011.5872[hep-th].


\bibitem{Gogberashvili2012}
    M. Gogberashvili, P. Midodashvili, and L. Midodashvili,
    Int. J. Mod. Phys. \textbf{D 21} (2012) 1250081,
    arXiv:1209.3815[hep-th].



\bibitem{dewolfe}
 O. DeWolfe, D. Z. Freedman, S. S. Gubser, and A. Karch,
    Phys. Rev. \textbf{D 62} (2000) 046008,
    arXiv:hep-th/9909134.


\end{thebibliography}
\end{document}